\documentclass[twocolumn,showpacs,preprintnumbers,amsmath,amssymb]{revtex4-1}

\usepackage{latexsym,verbatim}
\usepackage{amssymb}
\usepackage{amsfonts}
\usepackage{amsmath,color}
\usepackage{graphicx} 
\usepackage{bm}
\usepackage[utf8]{inputenc}
\usepackage[T1]{fontenc}
\usepackage[toc,page]{appendix}
\usepackage{enumitem}

\usepackage{hyperref}

\hypersetup{
    colorlinks=true,
    linkcolor=red,
    citecolor=green,
    filecolor=magenta,      
    urlcolor=cyan,
    pdftitle={Einstein's elevator and the principle of equivalence},
    pdfpagemode=FullScreen,}

\def\ber{\begin{eqnarray}}
\def\eer{\end{eqnarray}}
\def\beq{\begin{equation}}
\def\eeq{\end{equation}}

\def\ed{\end{document}}

\makeatletter
\let\jnl@style=\rm
\def\ref@jnl#1{{\jnl@style#1}}

\def\aj{\ref@jnl{AJ}}                   % Astronomical Journal
\def\actaa{\ref@jnl{Acta Astron.}}      % Acta Astronomica
\def\araa{\ref@jnl{ARA\&A}}             % Annual Review of Astron and Astrophys
\def\apj{\ref@jnl{ApJ}}                 % Astrophysical Journal
\def\apjl{\ref@jnl{ApJ}}                % Astrophysical Journal, Letters
\def\apjs{\ref@jnl{ApJS}}               % Astrophysical Journal, Supplement
\def\ao{\ref@jnl{Appl.~Opt.}}           % Applied Optics
\def\apss{\ref@jnl{Ap\&SS}}             % Astrophysics and Space Science
\def\aap{\ref@jnl{A\&A}}                % Astronomy and Astrophysics
\def\aapr{\ref@jnl{A\&A~Rev.}}          % Astronomy and Astrophysics Reviews
\def\aaps{\ref@jnl{A\&AS}}              % Astronomy and Astrophysics, Supplement
\def\azh{\ref@jnl{AZh}}                 % Astronomicheskii Zhurnal
\def\baas{\ref@jnl{BAAS}}               % Bulletin of the AAS
\def\bac{\ref@jnl{Bull. astr. Inst. Czechosl.}}
\def\caa{\ref@jnl{Chinese Astron. Astrophys.}}
\def\cjaa{\ref@jnl{Chinese J. Astron. Astrophys.}}
\def\icarus{\ref@jnl{Icarus}}           % Icarus
\def\jcap{\ref@jnl{J. Cosmology Astropart. Phys.}}
\def\jrasc{\ref@jnl{JRASC}}             % Journal of the RAS of Canada
\def\memras{\ref@jnl{MmRAS}}            % Memoirs of the RAS
\def\mnras{\ref@jnl{MNRAS}}             % Monthly Notices of the RAS
\def\na{\ref@jnl{New A}}                % New Astronomy
\def\nar{\ref@jnl{New A Rev.}}          % New Astronomy Review
\def\pra{\ref@jnl{Phys.~Rev.~A}}        % Physical Review A: General Physics
\def\prb{\ref@jnl{Phys.~Rev.~B}}        % Physical Review B: Solid State
\def\prc{\ref@jnl{Phys.~Rev.~C}}        % Physical Review C
\def\prd{\ref@jnl{Phys.~Rev.~D}}        % Physical Review D
\def\pre{\ref@jnl{Phys.~Rev.~E}}        % Physical Review E
\def\prl{\ref@jnl{Phys.~Rev.~Lett.}}    % Physical Review Letters
\def\pasa{\ref@jnl{PASA}}               % Publications of the Astron. Soc. of Australia
\def\pasp{\ref@jnl{PASP}}               % Publications of the ASP
\def\pasj{\ref@jnl{PASJ}}               % Publications of the ASJ
\def\rmxaa{\ref@jnl{Rev. Mexicana Astron. Astrofis.}}%
\def\qjras{\ref@jnl{QJRAS}}             % Quarterly Journal of the RAS
\def\skytel{\ref@jnl{S\&T}}             % Sky and Telescope
\def\solphys{\ref@jnl{Sol.~Phys.}}      % Solar Physics
\def\sovast{\ref@jnl{Soviet~Ast.}}      % Soviet Astronomy
\def\ssr{\ref@jnl{Space~Sci.~Rev.}}     % Space Science Reviews
\def\zap{\ref@jnl{ZAp}}                 % Zeitschrift fuer Astrophysik
\def\nat{\ref@jnl{Nature}}              % Nature
\def\iaucirc{\ref@jnl{IAU~Circ.}}       % IAU Cirulars
\def\aplett{\ref@jnl{Astrophys.~Lett.}} % Astrophysics Letters
\def\apspr{\ref@jnl{Astrophys.~Space~Phys.~Res.}}
\def\bain{\ref@jnl{Bull.~Astron.~Inst.~Netherlands}}
\def\fcp{\ref@jnl{Fund.~Cosmic~Phys.}}  % Fundamental Cosmic Physics
\def\gca{\ref@jnl{Geochim.~Cosmochim.~Acta}}   % Geochimica Cosmochimica Acta
\def\grl{\ref@jnl{Geophys.~Res.~Lett.}} % Geophysics Research Letters
\def\jcp{\ref@jnl{J.~Chem.~Phys.}}      % Journal of Chemical Physics
\def\jgr{\ref@jnl{J.~Geophys.~Res.}}    % Journal of Geophysics Research
\def\jqsrt{\ref@jnl{J.~Quant.~Spec.~Radiat.~Transf.}}
\def\memsai{\ref@jnl{Mem.~Soc.~Astron.~Italiana}}
\def\nphysa{\ref@jnl{Nucl.~Phys.~A}}   % Nuclear Physics A
\def\physrep{\ref@jnl{Phys.~Rep.}}   % Physics Reports
\def\physscr{\ref@jnl{Phys.~Scr}}   % Physica Scripta
\def\planss{\ref@jnl{Planet.~Space~Sci.}}   % Planetary Space Science
\def\procspie{\ref@jnl{Proc.~SPIE}}   % Proceedings of the SPIE

\makeatother

\begin{document}

\author{Lisa Galvagni}
\email{lisa.galvagni@gmail.com}
\affiliation{Dipartimento di Matematica ``G.Peano'', Universit\`a degli studi di Torino, Via Carlo Alberto 10, 10123 Torino, Italy}

\author{Guido Magnano}
\email{guido.magnano@unito.it}
\affiliation{Dipartimento di Matematica ``G.Peano'', Universit\`a degli studi di Torino, Via Carlo Alberto 10, 10123 Torino, Italy}

\author{Matteo Luca Ruggiero}
\thanks{Corresponding Author}
\email{matteoluca.ruggiero@unito.it}
\affiliation{INFN - LNL , Viale dell'Universit\`a 2, 35020 Legnaro (PD), Italy}
\affiliation{Dipartimento di Matematica ``G.Peano'', Universit\`a degli studi di Torino, Via Carlo Alberto 10, 10123 Torino, Italy}

\date{\today}

\title{Einstein's elevator and the principle of equivalence}
\begin{abstract}
We outline here the design, execution, and educational outcomes of an  intervention inspired by Einstein's elevator thought experiment, intended to introduce secondary school students to the principle of equivalence, which is at the basis of the theory of General Relativity. We build an experimental version of Einstein's elevator, which simulated the effects of free-fall in an accelerated reference frame: a detailed description of the experimental apparatus and its construction is provided, highlighting the challenges and innovations in creating a simple yet functional setup using everyday materials.
\end{abstract}
\maketitle

%%------------------------Section-------------------------
\section{Introduction }\label{sec:intro}
%%------------------------Section-------------------------

What an elevator has to do with one of the deepest ideas in physics? Einstein's elevator is one of the most famous \textit{Gedankenexperimente} (``thought experiments'') used by the German physicist to investigate the new laws of physics, going beyond the limitations of real laboratories. Riding beams of light was the image which stimulated his reflections on the finiteness of the speed of light and led to the theory of special relativity.  Subsequently, Einstein noted that within a windowless elevator an observer would be unable  to distinguish whether the elevator is stationary in a gravitational field or accelerating upward at a constant rate. Then  the fundamental laws of physics must be identical in both scenarios: this idea, known as the ``principle of equivalence'' asserts that, on a local scale (inside the elevator), the effects of gravity are indistinguishable from those of acceleration in a gravity-free environment and, similarly, that an elevator in free fall is indistinguishable from a Newtonian inertial frame \cite{sabine}.  The principle of equivalence, which is based on the identity of inertial and gravitational mass,  lies at the heart of the theory of general relativity, and makes it possible to describe gravity as the geometry of spacetime \cite{rindler2006relativity}.  Actually, the equivalence of inertial and gravitational mass,  far from being a purely abstract concept, underpins phenomena we encounter daily. For instance, it enables amusement parks to create experiences of ‘zero gravity’ in free fall towers or roller coasters, turning theoretical physics into tangible thrills \cite{pendrill2014equivalence}. 

As Dr. Derek Muller highlighted on his YouTube channel: ``If we want to teach children about the world as we understand it, we must teach them modern science, our most complete knowledge to date. Scientific literacy is extremely important in our media-saturated world, and today’s scientific topics include things such as renewable energy and black holes''\cite{Veritasium}.  
General relativity is our most accurate model of gravity, having successfully passed numerous observational tests \cite{will2014confrontation}. However, along with quantum mechanics—the other cornerstone of modern physics—it is typically excluded from schools curricula, except for brief overviews in the final years of some secondary schools. These theories are significant not only for their revolutionary impact on physics but also for their practical applications, which profoundly influence daily life; however, they remain largely inaccessible to most people. Recent studies suggest that introducing modern physics at an early stage in education is both feasible and effective: the Einstein First project \cite{pitts2014exploratory,kaur2017teaching1,kaur2017teaching2,kaur2017teaching3,kaur2017gender,kaur2017evaluation,foppoli2018public,choudhary2018can,2021PhyEd..56e5031A,2022arXiv220300842K,2023arXiv230617344K,2024PhyEd..59f5008K} seeks to integrate modern physics concepts into learning sequences as early as primary school.
Italy, where our research has been carried out, makes no exception since general relativity is not part of  schools curricula, where only very few hours are dedicated to modern physics in some specific school tracks. 

This work describes a research study performed in the second-to-last year and last year of secondary school, with the purpose to introduce the basic ideas of general relativity, starting from the principle of equivalence: to make it accessible to high school students, we used an experimental apparatus to recreate Einstein's elevator thought experiment. This approach provides a hands-on, tangible perspective on the theory, effectively bridging abstract concepts with real-world applications. However, this task presents significant challenges: the experimental setup must be designed using inexpensive materials to ensure replicability in classrooms and, as we said before, we must present the theory in classes that have only been exposed to classical physics. Our teaching strategy also involves engaging students with their existing knowledge of Newtonian mechanics and highlighting its limitations. For this purpose, we used a famous excerpt from the book written by \citet{einstein1966evolution}, where an imaginary dialogue with a classical physicist takes place: this enables students to discover the equivalence principle themselves through inquiry-based learning and experimentation.  Our aim is to make the fundamental concepts of general relativity accessible to high school students, with the convincement that the guiding ideas which inspired the theory can be understood without mastering its complex mathematical framework.

The paper is organized as follows: in Section \ref{sec:project} we describe the development of the intervention, while in Section \ref{sec:exp} we focus on the structure of the lectures and their methodologies. Einstein's elevator is discussed in Section \ref{sec:Einstein}, both in terms of its didactic relevance and with regard to its experimental construction. The results are discussed in Section  \ref{sec:ReD} and the conclusions can be found in Section \ref{sec:concl}.

%%------------------------Section-------------------------
\section{The Intervention}\label{sec:project}

\subsection{Preliminary phase}

To design the intervention, we made a survey of the most used physics textbooks in Italian high schools, which are those written by \citet{cutnell2019fisica}, \citet{walker2012dalla}, \citet{amaldi2016dalla}, to understand how they introduce general relativity and which are the most relevant topics covered. We made a list of these topics and shared it with 29 high school teachers who are involved in training programs developed at the University of Turin. Then, we asked them to choose the most important topics for an introduction to general relativity: they turn out to be the principle of equivalence, the principle of general covariance, the propagation of light in curved spacetime and the experimental tests of general relativity. Among them, we chose  the principle of equivalence to introduce Einstein's theory: due to very limited time at our disposal, our choice was dictated both by the purpose to explain an idea that lies at the core of general relativity and by the possibility to build an actual experiment to reproduce Einstein's thought experiment.

\subsection{Time allocated and structure of the intervention}
The research study involved 115 students from five classes in two high schools of classical studies in Turin, Italy. In particular, two of the participant classes were in the second-to-last year and three in the last year. Of them, only one had already studied special relativity. The intervention, which lasted for 4 hours and was conducted in April and May (a particularly critical period for the last year classes due to the upcoming final exams), was divided into two lessons of 2 hours each, preceded by a test that students completed independently at home. The test was based on an excerpt taken from the book written by \citet{einstein1966evolution}, the so-called ``Dialogue with the Classical Physicist'',  on the meaning of inertial reference frames. 

The first lesson focused on the equivalence principle: the first hour provided a theoretical introduction, emphasizing the doubts that Einstein raised regarding Newton's theory. The second hour was devoted to analyzing and resolving these doubts through the elevator experiment.

The second lesson focused on the meaning of gravity as geometry of spacetime and on the experimental tests of general relativity. 

At the end of the four hours of lessons, a final test was provided to teachers to administer to the students, with the aim of assessing their understanding of the topics covered. The test was divided into two parts: the first part consisted of four multiple choice questions on the theory presented during the lessons; the second part consisted of three open-ended questions based on three short scientific texts. To answer these latter questions, students had to analyze and understand the proposed texts, apply the concepts of general relativity they had learned, and provide well-reasoned responses referencing Einstein's theory, as explained during the lessons.

%%------------------------Section-------------------------
\section{Lecture's structure and applied methodologies}\label{sec:exp}
\subsection{Initial test}

Before starting the first activity, students were asked to read the already cited excerpt  (see Appendix \ref{app:dialogue}) taken by the book written by \citet{einstein1966evolution} and answer five multiple-choice questions, which can be found in Appendix  \ref{app:initialtest}. The last question also required a brief explanation of the chosen answer. The responses collected allowed for an assessment of the student's level of knowledge before beginning the learning process.

This test had multiple purposes, the primary one being to compare students' knowledge before and after the in-class activities. However, it was not the only objective. Since four of the five participating classes had never studied Einsteinian physics, it was necessary to introduce the topic using concepts already familiar to them. Given their foundation in Newtonian physics, we assigned a reading in which Einstein highlights some of the shortcomings of classical physics.

The first four questions in the initial test were designed to encourage students to reflect on the concept of inertial reference frames and the force of gravity, two key themes in Einstein’s theory, which are completely redefined with respect to their meaning in classical physics.   The last question, more challenging, asked students to align themselves with Einstein, classical physics, both, or neither, providing justification for their choice. The primary purpose of this question was not so much to elicit the “correct” answer but to encourage students to present a well-reasoned and articulated explanation.  An  analysis of the answers given by the students will be presented in Section \ref{Initial_test_res}.

\subsection{Theoretical introduction}

The first classroom session began with a presentation and a brief introduction to the work that would be carried out. This was followed by a collective discussion to address the issues raised in the text presented in the initial questionnaire and to ensure that all students clearly understood these issues. Particular attention was given to highlighting the problems related to inertial reference frames and gravity as a force in Newton's theory. A historical introduction was then provided  to illustrate Newton's theory and the conceptual difficulties emphasized by Einstein. For instance, to stimulate students' reflections on the gravitational interactions, we used the question: “The distance between the Sun and the Earth is 150 million kilometers. What kind of force acts over such a distance?” After introducing the doubts raised by Einstein, students were shown the famous statement made by Einstein that shook the world of physics at the time: “Gravity is not a force.”

\subsection{Introduction to Einstein's elevator}

Once students clearly understood the issues in Newton's theory highlighted by Einstein (e.g., What is an inertial reference frame? What is gravity if it is not a force?), we introduced Einstein's elevator. The mental experiment conceived by Einstein was first explained, and students were then asked how a person inside the elevator would perceive gravity. It was repeatedly emphasized that the elevator serves solely to remove any external reference points for the person inside and that the air resistance acting on the elevator during free fall should not be considered.

Afterward, students were shown the experimental setup that would be used to address Einstein's question. However, before observing the experiment, they were asked to formulate hypotheses about its outcome. The hypotheses formulated by the students were threefold: \\
\begin{enumerate}
\item  The person inside the elevator crashes into the ceiling.
\item   The person inside the elevator feels heavier.
\item The least popular hypothesis: the person inside the elevator rises and floats.
\end{enumerate}
As can be clearly observed, the teaching methodology employed is Inquiry-Based Learning (IBL). A method in which effort is therefore followed in order to develop the creative thinking of pupils at the expense of a drill and memorizing. Furthermore, it also develops the skills to solve unknown situations that a pupil will face later in his/her life. Very often people have to apply what  they have learnt in new situations and, to do that, they need to control their basic thinking and other general cognitive skills which create the essence of the individual competence for the problem solving \cite{dostal2015definition}. Inquiry experiences can provide valuable opportunities for students to improve their understanding of both science content and scientific practices \cite{edelson1999addressing}.

\subsection{Formulation of the equivalence principle}

During the experiment it was observed that at the moment of free fall, the reading of the dynamometer is zero: this  explains why the object or person inside the elevator experiences the sensation of “floating,” as if in a state of weightlessness. As a matter of fact,  the weight recorded by the dynamometer is zero. After disproving the first two hypotheses formulated by the students, a collective discussion was initiated to analyze the observations from a theoretical perspective. The sensation of weightlessness and floating inside the elevator, along with the fact that the spring dynamometer registers {zero weight}, does not guarantee being sufficiently far from a massive body or free from gravitational influences. Building on this reasoning, the principle of equivalence was introduced, encouraging students to articulate it spontaneously rather than presenting it as a “predefined and ready-made law.” To conclude the first two hours of the lesson in an engaging way and to set the stage for the next session, ensuring better assimilation of the concepts, a video was shown featuring the Italian astronaut Samantha Cristoforetti aboard the International Space Station, where she explains how mass and gravity work in space \cite{cristoforetti}.

%%------------------------Section-------------------------
\section{Einstein's elevator }\label{sec:Einstein}

\subsection{Projecting the elevator} \label{ssec:projelev}

Einstein's elevator is a thought experiment devised by Einstein to explain the equivalence principle. As we said, Einstein suggested that for an observer in an elevator uniformly accelerating in empty space objects fall as if they were under the influence of Earth's gravitational force when the elevator is stationary. The two scenarios are indistinguishable. Similarly, it is not possible to discriminate between an elevator in free fall within Earth's gravitational field and one at rest in empty space \cite{resnick1991introduction}. 
Our aim in \textit{building}  Einstein's elevator  was to allow students to "experience firsthand" the experiment proposed by Einstein: not to assume its validity without evidence, but to demonstrate it physically. The elevator described by Einstein was meant to fall continuously in free fall without stopping, so that the events inside could be clearly observed. Of course, it is not physically feasible to replicate such motion on Earth: however, the elevator should fall from a sufficient height to make the outcomes of the experiment observable. Additionally, the elevator needed to be sturdy enough to withstand the impact with the ground upon release. It should have included something to simulate a person inside, along with an instrument to measure any variations in weight during the fall. Furthermore, the interior of the elevator needed to be filmed so that the events occurring during the descent could be reviewed later.

Regarding the height from which the elevator was supposed to fall (undoubtedly the most delicate issue to resolve!) it was essential to design a system that could be easily replicated in all classrooms and conveniently transported, since the intervention took place in diverse schools. Initially, we considered a metal structure made from four 50 cm pieces that could be assembled to reach a total height of 2 m. However, this design proved unsuitable because it would have been difficult to construct and would not support the weight of the elevator during its fall. We then looked for an existing object approximately two meters tall and thought of classroom doors, which are present in all schools and sturdy enough to anchor the elevator using clamps and ropes.

To simulate the elevator, we decided to use a wooden box and cushion its landing with foam placed on the floor near the door. Inside the box, to represent a person, we would place a {200 g mass} attached to a dynamometer, which would be fixed to the ceiling of the box. This setup would allow us to measure fluctuations in weight during the fall. To record the experiment, we planned to design a structure (still undefined at the project's inception) to hold a smartphone. The box would be suspended by a system of ropes and pulleys that would lift it to the top of the door before releasing it to fall freely to the ground. This arrangement would enable students to conduct the experiment independently and safely.

\subsection{Building the elevator} \label{ssec:buildelev}

{This section outlines the steps involved in building the experimental setup, accompanied by images to enhance clarity. Additional photographs can be found in Appendix \ref{app:fig} to better illustrate the completed setup, and a video demonstrating both the apparatus and its operation is available here \cite{video}.}

\begin{figure}[h] 
\centering 
\includegraphics[width=1.0\columnwidth]{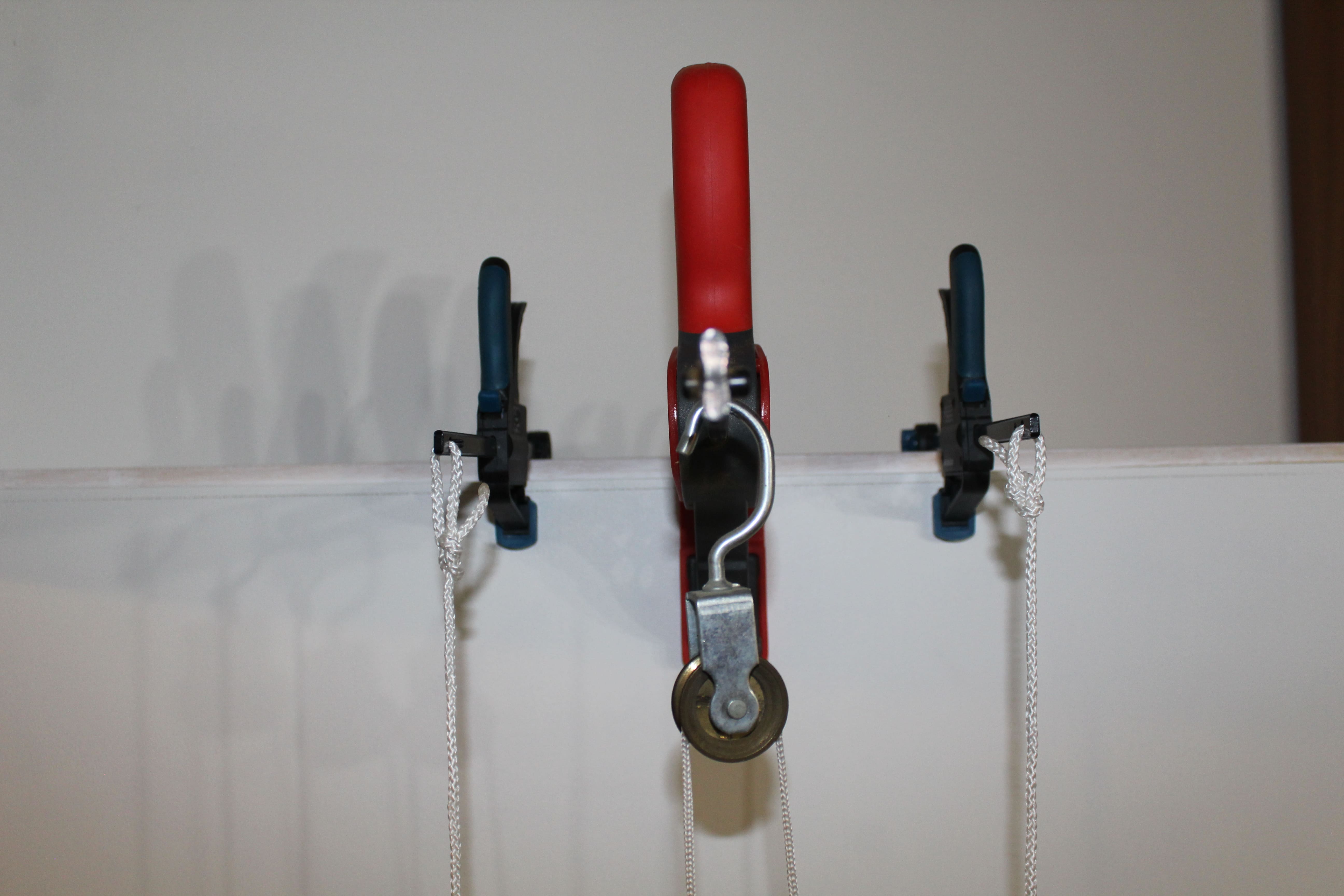} 
\caption{Setup of the three clamps: one larger central clamp and two smaller side clamps. The central clamp secures the pulley that supports the main rope, which is responsible for driving the elevator's movement. The side clamps anchor the guide ropes using simple knots, ensuring that the ropes pass smoothly through metal rings attached to the box's side walls. This arrangement maintains proper alignment and prevents the ropes from bending.}\label{figura:morsetti1}
\end{figure}

\begin{figure}[h] 
\centering 
\includegraphics[width=1.0\columnwidth]{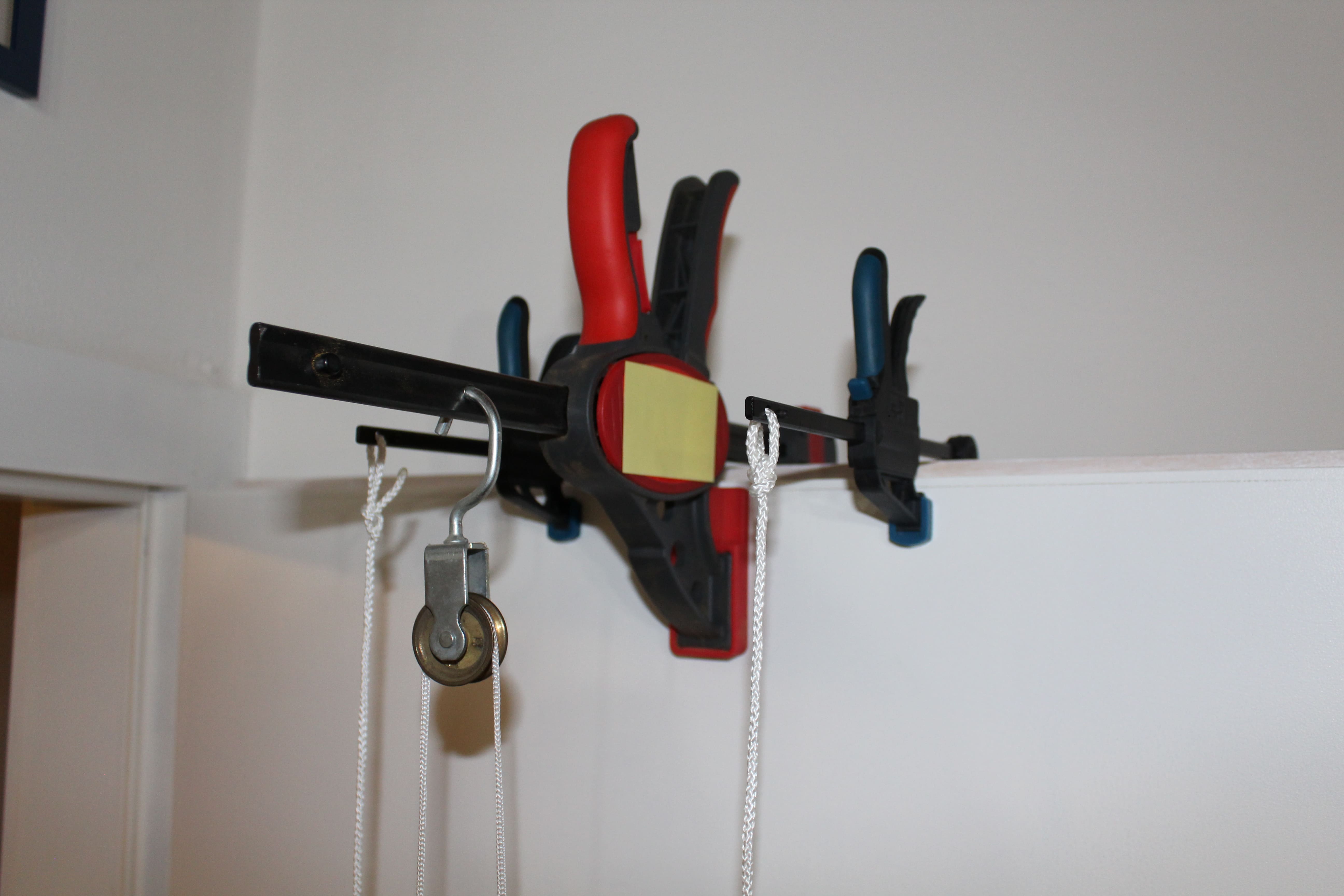} 
\caption{Side view of the three clumps setup.}\label{figura:morsetti2}
\end{figure}

To securely anchor the entire experimental apparatus to the classroom door, a system comprising three clamps was designed FIG. \ref{figura:morsetti1}. The central clamp, with a maximum load capacity of 68 kg and a maximum opening of 25 cm, is intended to support the weight of the elevator during its descent. Two additional clamps, positioned laterally to the central one and each with a maximum load capacity of 23 kg and a maximum opening of 25 cm, are used to hold the cables that serve as guides for the elevator. These lateral clamps ensure stability and prevent the elevator from rotating. A pulley was mounted on the central clamp, through which the cable supporting the elevator was threaded FIG. \ref{figura:morsetti2}. 

\begin{figure}[h] 
\centering 
\includegraphics[width=1.0\columnwidth]{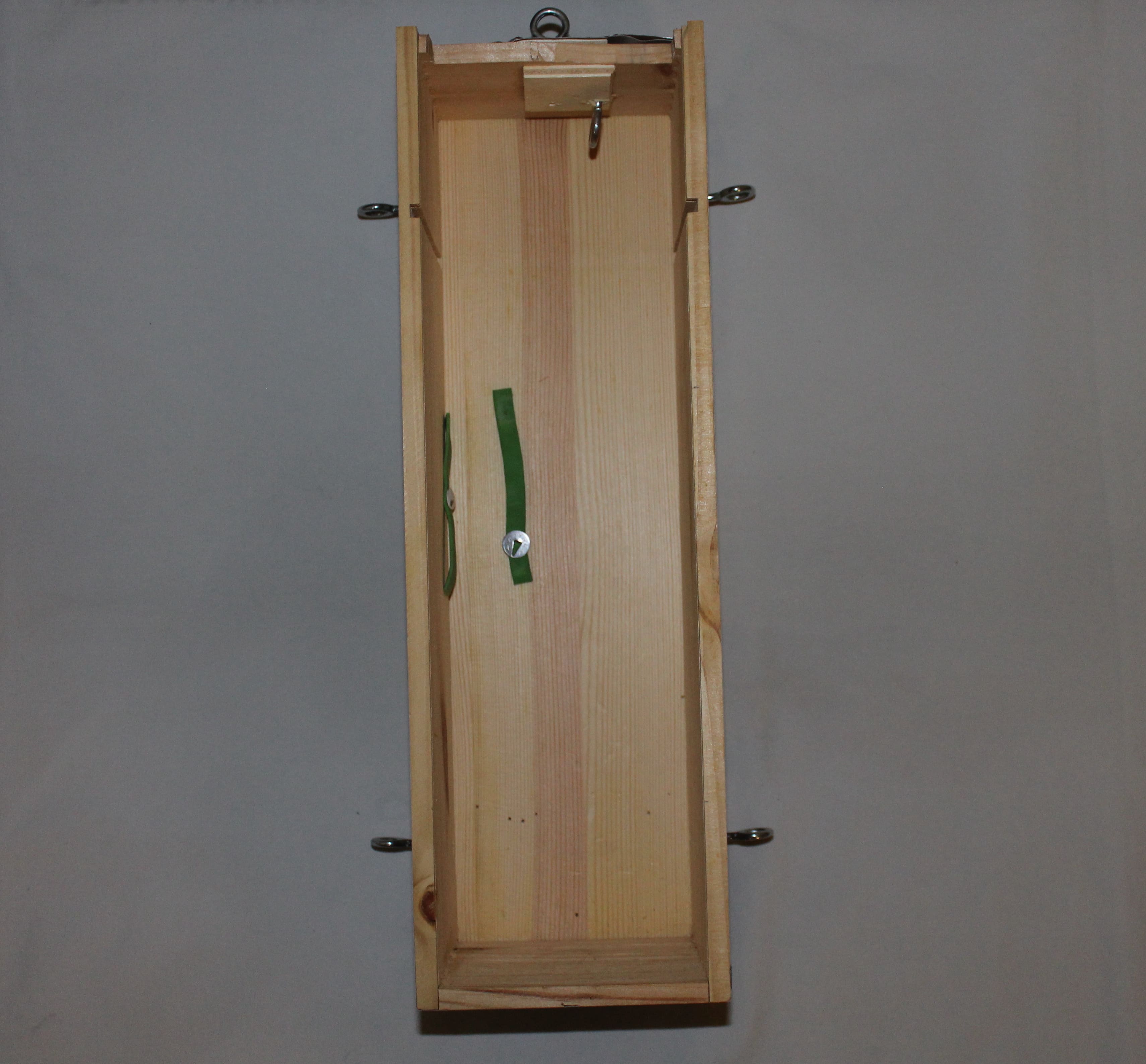} 
\caption{The box (12 cm x 12 cm x 38 cm) without its lid, showcasing the six metal rings: four rings on the sides, through which the two guide ropes must pass; two rings on the upper wall of the box: one inside for attaching the dynamometer and one outside for securing the main rope that moves the box.}\label{figura:scatola1}
\end{figure}

\begin{figure}[h] 
\centering 
\includegraphics[width=1.0\columnwidth]{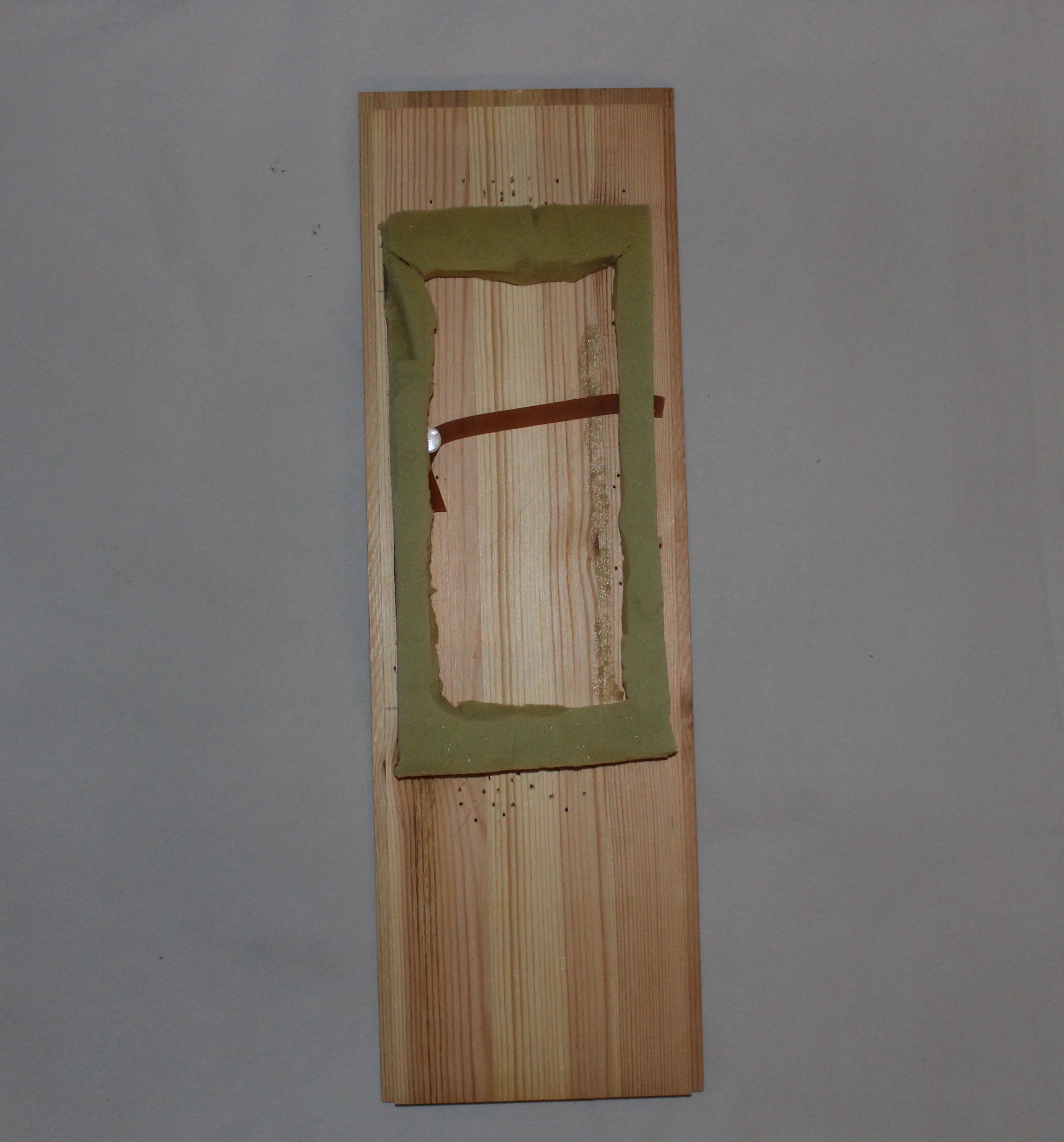} 
\caption{The mechanism designed to secure the smartphone to the lid of the box.}\label{figura:coperchio1}
\end{figure}

The elevator was simulated using a box originally designed to hold a wine bottle. Six holes were created in the box: two on the top surface to secure a metal ring for supporting the dynamometer and the central cable, and four on the lateral walls, arranged in pairs. These lateral holes were used to accommodate metal rings, allowing cables connected to the two lateral clamps - serving as elevator guiding tracks - to pass through FIG. \ref{figura:scatola1}. The lid of the box features a sliding interlocking mechanism, eliminating the need for additional equipment to keep it closed during descent. A cut was made along the top surface of the box to ensure the lid could close seamlessly, even with the smartphone attached. A dedicated support structure for the smartphone was built on the lid. A rectangular foam frame, which matches the dimensions of the smartphone, was cut and affixed to the lid. To further secure the smartphone, an elastic cord was attached outside the foam frame, anchored with two push pins placed at the midpoint of the longer sides of the rectangle FIG. \ref{figura:coperchio1}.

\begin{figure}[h] 
\centering 
\includegraphics[width=1.0\columnwidth, angle=90,origin=c]{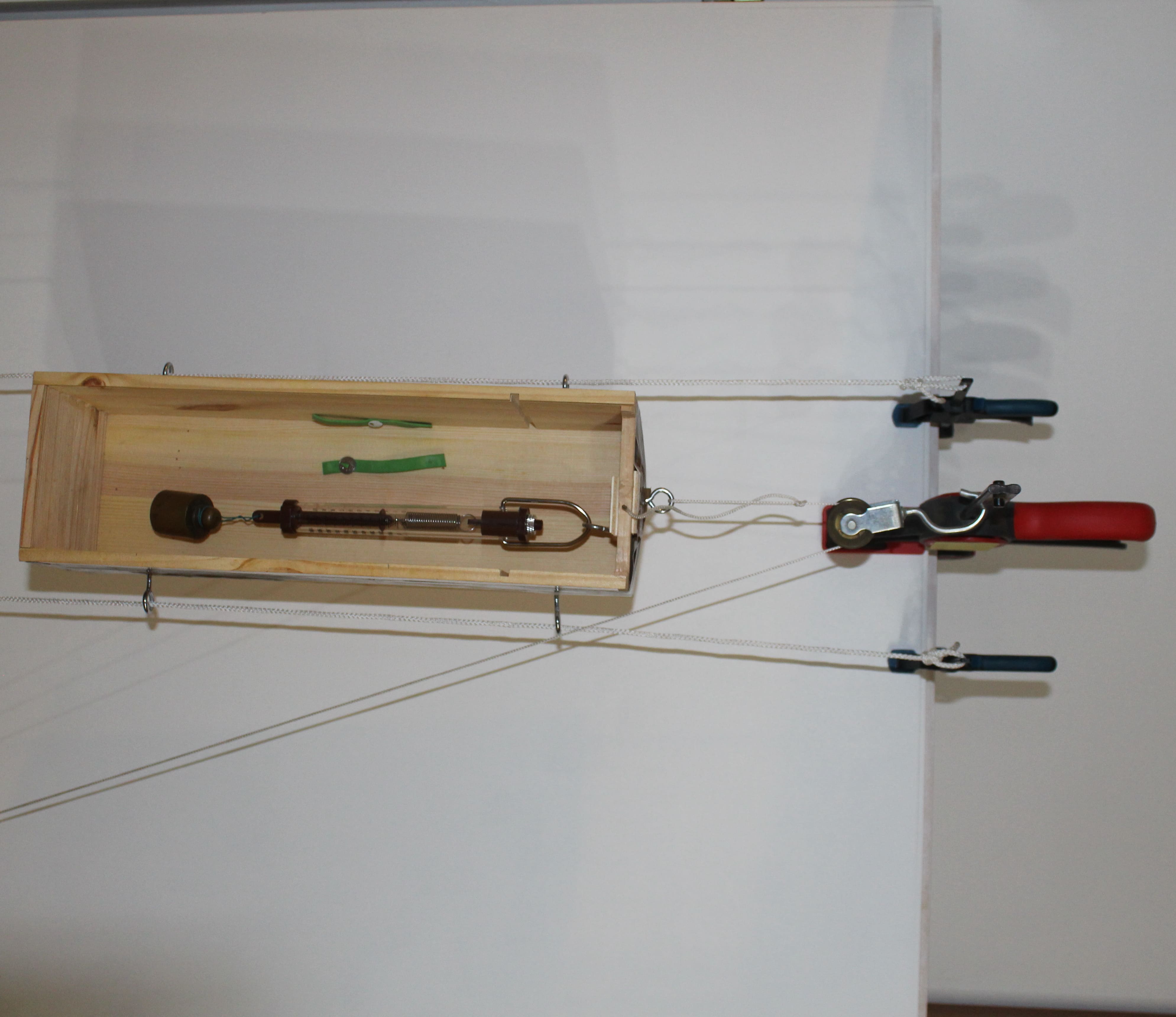} 
\caption{The picture demonstrates how the box is held by the ropes linked to the clamps, as well as the positioning of the dynamometer and the weight inside.}\label{figura:scatola2} 
\end{figure}

\begin{figure}[h] 
\centering 
\includegraphics[width=1.0\columnwidth]{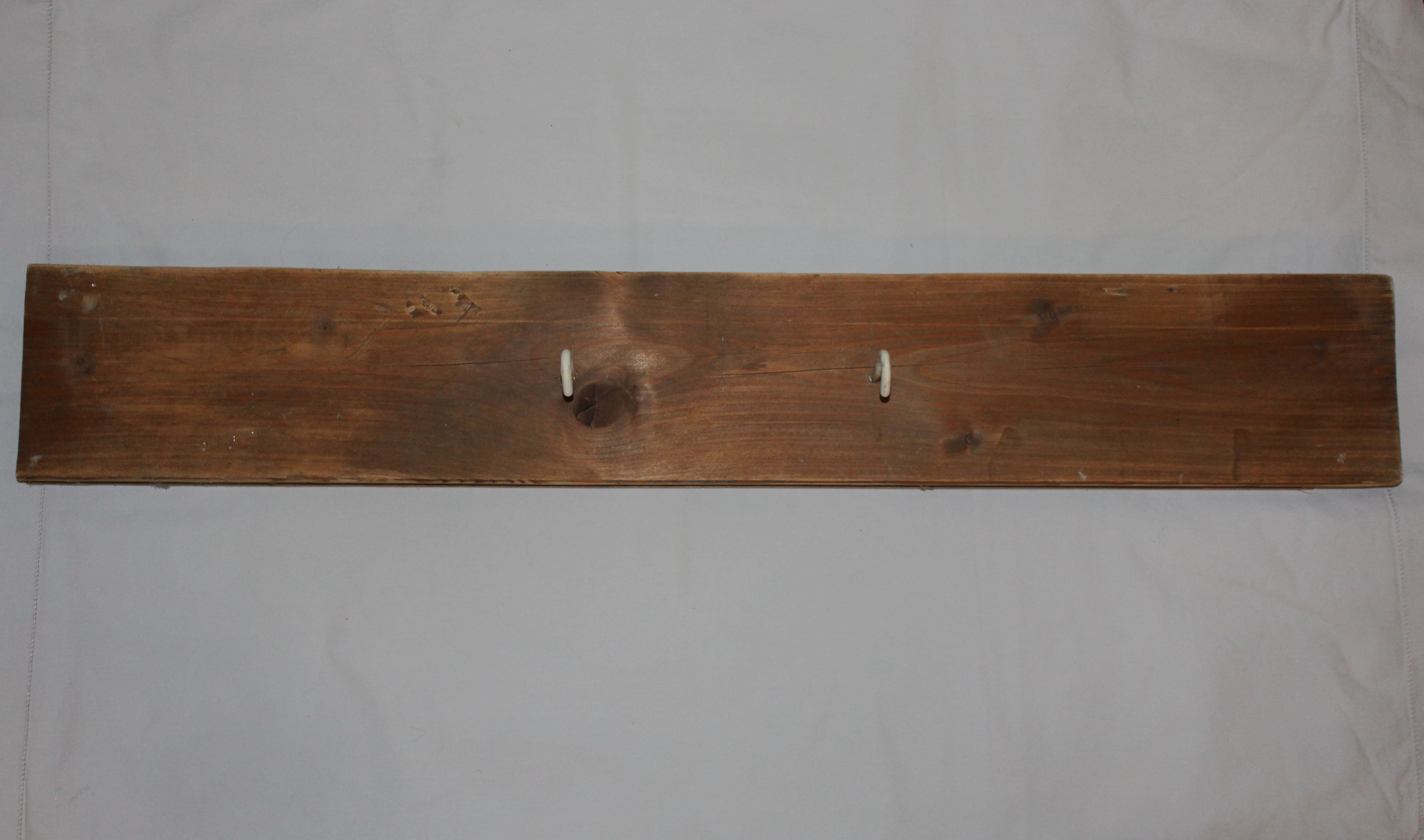} 
\caption{Wooden plank placed on the floor by the door to cushion the box impact.}\label{figura:base1}
\end{figure}

\begin{figure}[h] 
\centering 
\includegraphics[width=1.0\columnwidth]{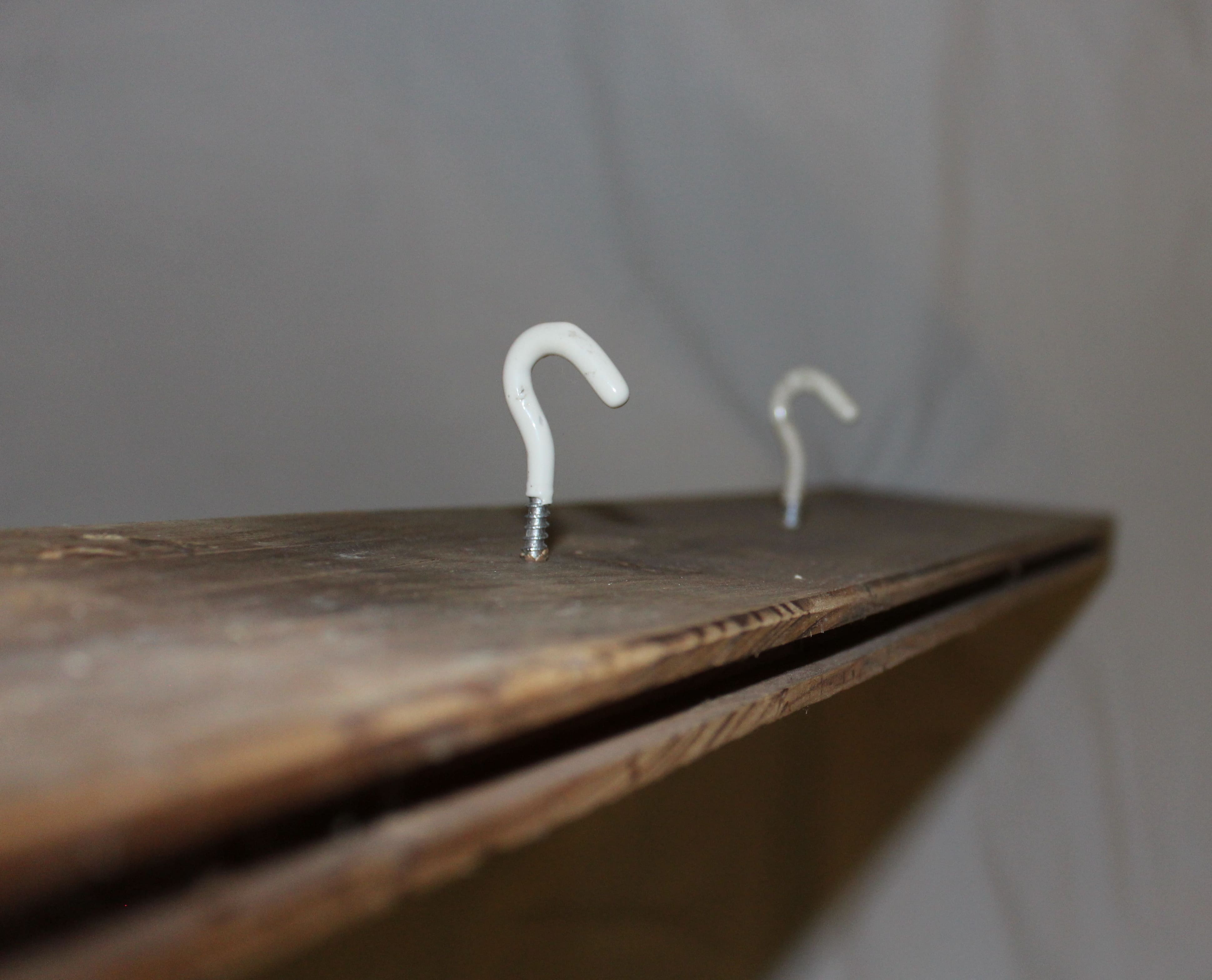} 
\caption{The hooks supporting the two guide ropes in the wooden plank.}\label{figura:base2}
\end{figure}

\begin{figure}[h] 
\centering 
\includegraphics[width=1.0\columnwidth]{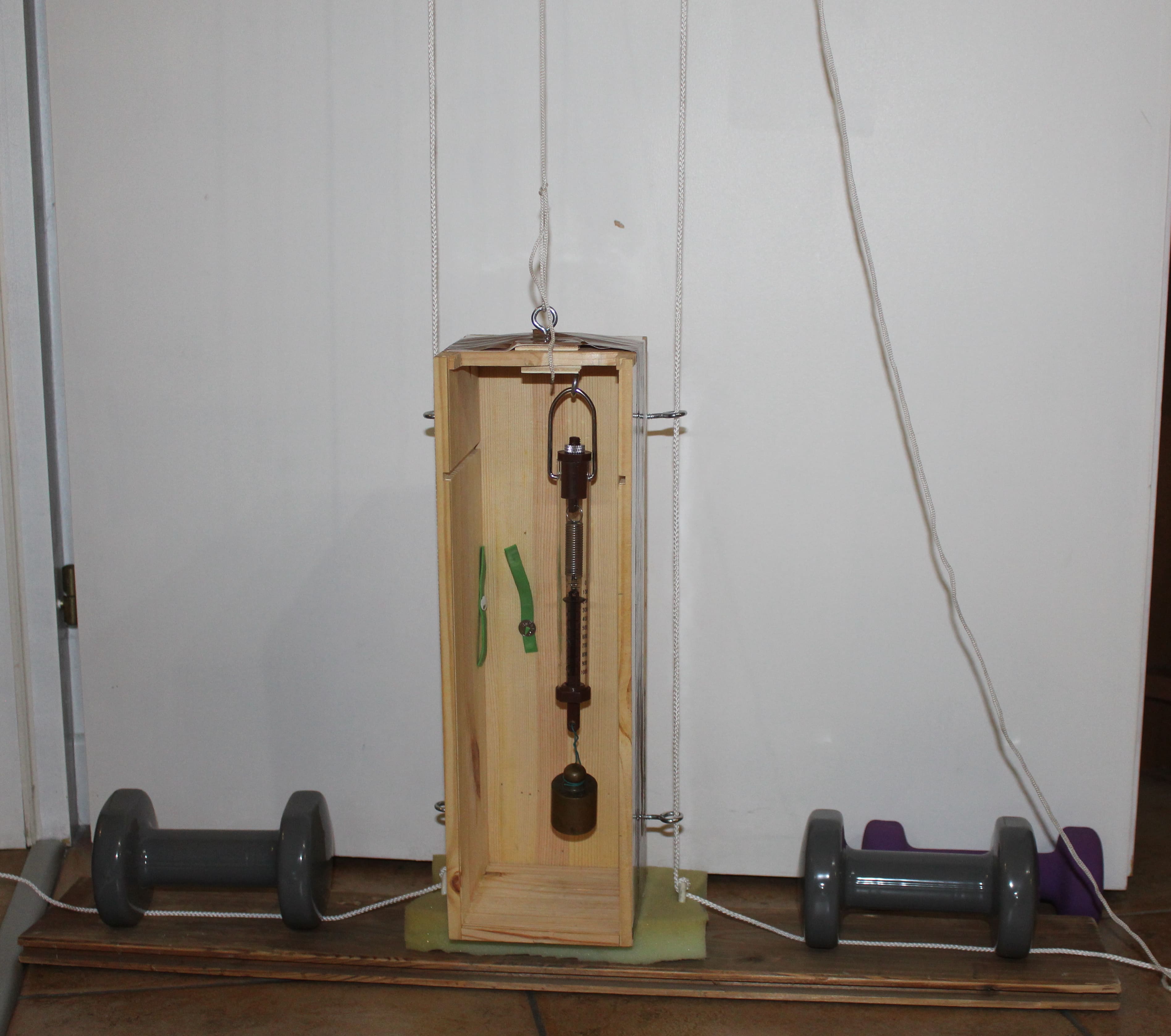} 
\caption{The elevator on the wooden base, with the weights securing the ropes and the foam placed underneath the box visible.}\label{figura:base3} 
\end{figure}

As we said, to simulate a person inside the elevator, a {mass of 200 g}  was suspended from a dynamometer with a measurement range of up to 1 kg. The dynamometer was attached to a hook on the upper wall of the box, ensuring that the weight remained suspended without contacting the lower surface FIG. \ref{figura:scatola2}. At the base of the door, a wooden plank measuring 75 cm x 12 cm x 1.5 cm was installed FIG. \ref{figura:base1}. Two hooks were affixed to the plank at intervals that corresponded to the width of the box FIG. \ref{figura:base2}. This setup served two purposes: first, to cushion the box's impact by placing foam material between the clamps, and second, to secure the ropes that act as guides for the elevator. To stabilize the ropes and keep the plank in position, weights were added to both ends of the plank FIG. \ref{figura:base3}. 

\subsection{Practice tests and modifications}
After constructing the experimental apparatus, we began practical tests to evaluate its functionality before presenting it in educational settings. The initial trials revealed several issues. The smartphone camera, positioned too close to the dynamometer, was unable to effectively capture its oscillations. Additionally, the interior of the box lacked sufficient illumination, resulting in poor visibility of the experiment in the recorded footage.

To address these issues, we implemented two modifications. First, we increased the screen framing to 0.5x to allow for better observation of the dynamometer’s oscillations. Second, we activated the smartphone’s flashlight to illuminate the interior of the box during the fall. These adjustments led to the successful acquisition of a valid recording, enabling clear visualization of the dynamometer's oscillations. The second series of tests produced satisfactory footage. By analyzing the video in slow motion, we observed weight fluctuations during free fall, with the dynamometer briefly stabilizing at the 0 N mark before the apparatus reached the ground.

\begin{figure}[h] 
\centering 
\includegraphics[width=1.0\columnwidth, angle=90,origin=c  ]{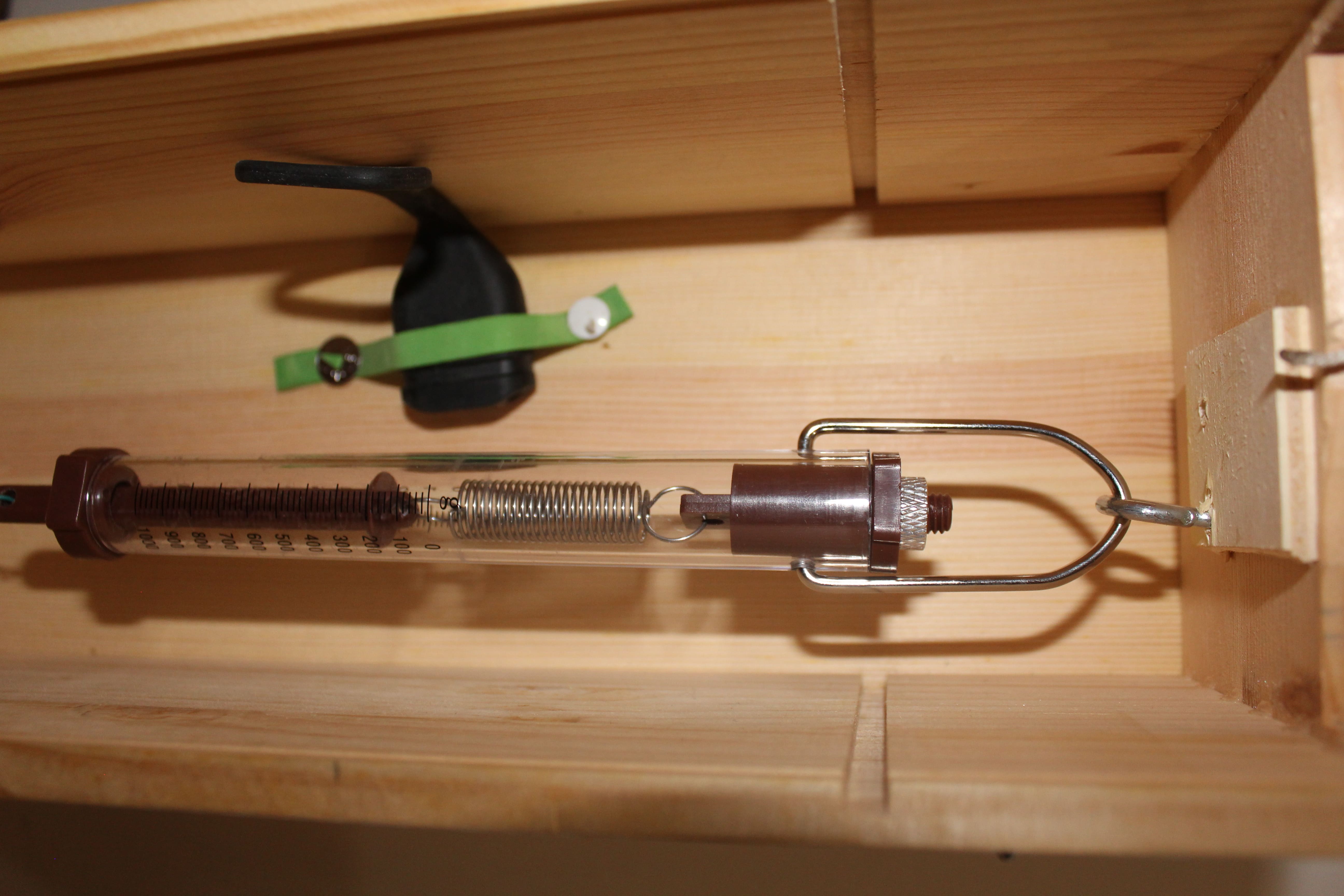} 
\caption{Mechanism to secure the light to the box using a rubber band and two push pins.}\label{figura:luce1} 
\end{figure}

Further refinements were necessary to improve the quality of the recordings. The direct glare from the smartphone flashlight caused reflections that obscured the dynamometer's scale. To resolve this, we installed a separate white light source—commonly used for bicycle illumination—inside the box to provide even lighting without reflections FIG. \ref{figura:luce1}.
\vspace{3mm}
\\
The third series of tests yielded the best results. During these trials, we noted that the box needed to be released from a stationary position to minimize disturbances. By ensuring a stable initial setup, we significantly reduced the oscillations of the dynamometer caused by movements of the box and pulley during free fall.

%%------------------------Section-------------------------
\section{Discussion of the Results}\label{sec:ReD}

\subsection{The initial test}\label{Initial_test_res}

As for the initial test, 80\% of the students answered question 1 correctly, 85\% answered question 2 correctly, 17\% answered question 3 correctly, and 76\% answered question 4 correctly. Question 3 caused the most confusion because it directly referred to the effect of gravitational force on inertial reference frames.
During the first classroom lesson, students were asked to verbally explain their reasoning for this question. Most of them stated that they found it difficult due to their understandings of the following facts:\\
a) Gravitational force is a constant presence on Earth.\\
b) No external forces should act on an inertial reference system.

They struggled to "combine" these two aspects of physics, and in their attempt to do so, the second concept prevailed. In order to better understand the reasoning behind the students' answers, during the first lesson they were also asked to explain how they answered and justified the final question. The latter required them to decide whether they agreed with Einstein, the classical physicist, neither, or both, and to justify their choice.

An analysis of their responses reveals that students generally place great trust in the classical physics  they learn in school. However, despite this trust, most did not choose the classical physicist in their answer to the final question. Many stated that they believed both Einstein and the classical physicist were correct, while others simply said, "I don’t feel comfortable saying Einstein is wrong." They clearly recognize Einstein as one of the greatest minds in history, although many lack a detailed understanding of why he holds that status. This is likely due to the fact that their education in physics has primarily focused on concepts from before the 20th century. Among the 115 students involved,  38 chose to side exclusively with Einstein. While some provided well-reasoned justifications, others expressed sentiments such as, "Einstein cannot be wrong".

\subsection{During the lesson}
It is interesting to point out that the students' attention grew significantly when the assertions of classical physics were questioned through sound reasoning. The entire class was captivated when Einstein's doubts were presented, especially when they encountered the scientist's statement: "Gravity is not a force."

During the segment dedicated to Einstein's elevator experiment, most students showed eagerness and a willingness to make the experimental apparatus work. They were curious to see what would actually happen and to confirm or refute their hypotheses. Students in each class participated in the experiment, from assembling the apparatus to dismantling it. They wanted to repeat the experiment multiple times to ensure its success. This process allowed them to understand the challenges scientists face daily when attempting to verify or disprove their hypotheses. When we slowed down the video captured with the smartphone inside the box, the students noticed that not all the trials were successful. In some tests, accidental errors occurred—such as the elevator oscillating too much, which caused the dynamometer to move erratically; the camera failing to focus on the dynamometer; the light inside the box turning off during the fall; the dynamometer detaching from the hook at the start of the fall; or the smartphone dislodging from its mount as the elevator began to drop.

Despite these setbacks, in each class, we managed to capture at least one recording that demonstrated what the experiment aimed to prove Einstein's thought experiment hypothesis. All the students were astonished by the outcome, as only one or two students per class had initially predicted the correct result of the experiment. After the experiment, the students themselves wanted to seek a theoretical explanation for what had happened. This confirmed their interest in the topic, even though it concerned a subject most of them generally dislike.

Gaining the students' attention during the lesson was crucial for making the session as interactive as planned. While engaging language was important, the key element was Einstein's simple yet profound statement: "Gravity is not a force." This statement, both clear and thought-provoking, resonated with the students, who were used to thinking of gravity as a force. Their recognition of Einstein as one of the most significant scientists in history added to their intrigue. This combination of confusion and curiosity encouraged them to seek answers to the questions that emerged.

\subsection{{Final test}}

{The file containing the questions for the final test (Appendix \ref{app:finaltest}) was emailed to the teachers, who subsequently shared it with their students. The test duration was approximately one hour per class, although some classes completed it in as little as 45 minutes. It was administered in paper format, and no personal data from the students was collected. Out of a total of 115 students, 109 participated and answered the final questions.
\vspace{3mm}
\\
Due to school scheduling, the final test was administered by the teachers during regular class hours. As a result, a few students who had not attended the lessons also took the test. Although the exact number of these students was not recorded, it is known that it does not exceed five. Their responses were included in the final sample; however, they do not significantly impact the overall results.
\vspace{3mm}
\\
Regarding the analysis of the students' responses to the final test, most of them provided the correct answers to the first four multiple-choice questions. A detailed analysis of the results shows that 90\% of the students answered the frst question correctly, 72\% answered the second, 73\% answered the third, and 77\% answered the fourth.
\vspace{3mm}
\\
In contrast, the open-ended questions proved to be more challenging and elicited more doubts and uncertainties. This type of question is inherently more complex and requires students to engage in deeper reasoning and demonstrate a greater understanding of specific concepts. The three texts referenced for the open-ended questions were: a document about the Gaia mission from the Italian Space Agency’s website \cite{ASI_Agenzia_Spaziale_Italiana} (Appendix \ref{app:gaia}), a text discussing the photon sphere of black holes \cite{Sfera_di_fotoni} (Appendix \ref{app:bh}) and an explanation of how GPS works \cite{GPS} (Appendix \ref{app:gps}).
\vspace{3mm}
\\
To analyze the results, we categorized the responses to the three open-ended questions as follows:
\begin{enumerate}[label=(\Alph*)]
\item Responses in which students referenced Einstein's theory and provided reasoning based on the new concepts presented during the four hours of lessons.
\item Responses in which students remained aligned with the theories and concepts of classical physics, offering arguments based on that framework.
\item Responses in which students did not reference any specific theory and provided incomplete answers with weak or often invalid reasoning.
\end{enumerate}
Our primary focus is to determine how many students transitioned from classical physics to Einstein's theory, indicating their understanding of the latter's propositions.
The data reveals that 44\% of the students fell into Category (A), 22\% into Category (B), and 34\% into Category (C).}

%%------------------------Section-------------------------
\section{Conclusions }\label{sec:concl}

In this paper we discussed a study was conducted in the final two years of Italian secondary school to introduce students to the fundamental concepts of general relativity, beginning with the principle of equivalence. To make the topic accessible, we used an experimental apparatus to replicate Einstein's elevator thought experiment.

According to our findings and within the limits of the sample considered, the results obtained are encouraging. In particular, the final test provided positive indications; however, due to unmodifiable conditions in that context (such as the anonymity constraints of the responses), an accurate quantitative evaluation was not possible. We aim to achieve this through further experiments in the future, potentially involving a larger sample of students.

A key feature of the intervention was the construction of experimental apparatus: in this respect, there is certainly room for improvement. To enhance the quality, it would be advisable to use a different internal camera rather than a smartphone, which would allow for better framing of the dynamometer and result in a higher-quality video. Additionally, designing a more stable attachment for the dynamometer, constructing a sturdier box, and equipping it with a consistent light source that remains on during the fall would be beneficial. To ensure proper visualization of the experiment, the box should ideally be dropped from a height greater than two meters; an additional meter would suffice. This could potentially be achieved by designing hooks attached to the ceiling and employing a braking system that is more effective than simply using 10 cm of foam rubber. A different braking method could be advantageous for preserving the integrity of the box, especially since significant damage could occur if it were dropped from a greater height.

Analyzing the strengths of this experiment, we can affirm that it is an experimental setup accessible to everyone. It was constructed with simple, readily available, and inexpensive materials, many of which can be found in most households, thus eliminating the need for new purchases. Although the execution of the experiment was not flawless, its outcome was satisfactory and met its intended purpose. Even though improvements can always be done, as with any experimental setups, we can consider the work accomplished a success, particularly because the experiment was effectively well conducted in all five classes involved.

Students were actively involved in the experiment, starting from the assembly of the experimental apparatus. With more time, it could be beneficial to project the apparatus together with the students, involving them from the outset to foster a sense of responsibility for the success or failure of the experiment. The teachers involved were enthusiastic about the work done, with some expressing a desire to propose similar lessons and experiments in future classes. On a practical level, the students particularly enjoyed the experimental phase. They were able to actively participate in the lesson and interact directly with the apparatus, which was specifically designed for them.

Based on the experience gained during this study, we can conclude that  a good approach for introducing general relativity at the secondary school level is through interactive lessons. We emphasize that most of the students involved did not possess a background in special relativity. 
Students should be fully engaged, with their questions stemming from their curiosity and motivation to seek answers. Lessons should allow for active participation, encourage students to ask questions, and, importantly, assure them that making mistakes is part of the learning process.  Emphasizing practical applications alongside theoretical explanations helps reinforce understanding. Hands-on experimentation allows students to grasp theoretical concepts more effectively.

In conclusion, our approach seems to suggest a possible way to introduce the basic ideas of modern physics which, as such, can be successfully understood by the students without the risk to obscure them behind a complex mathematical framework. The latter, of course, can be explored subsequently, but  it requires specific skills and competence that are not accessible to most of high school students.

\noindent \textbf{Ethical Statement:} Data used in this research were collected during an undergraduate thesis project. Pupils involved were informed about the treatment of data for research purposes.

%%------------------------Section-------------------------

\bibliography{ref}

%%--------------------------------------------------------------

\appendix

\section{Initial Test} \label{app:initialtest}

\begin{itemize}
\item \textbf{1.}
\textit{According to the classical physicist, an inertial reference system is an ideal concept useful for describing certain situations from a physical perspective. Therefore, we can say that:}
\vspace{3mm}
\\
A. On Earth, it is always possible to isolate a system to make it inertial.
\vspace{1mm}
\\
B. By eliminating only the Earth's rotation, it is possible to observe an inertial reference system on it.
\vspace{1mm}
\\
C. The Sun represents an inertial reference system.
\vspace{1mm}
\\
D. On Earth, it is not possible to find inertial reference systems, as there will always be at least one force acting on them.
\item \textbf{2.}
\textit{According to Einstein, who is the one asking the questions in the dialogue, the laws of classical physics are valid, but there is no framework to refer them to. What does Einstein mean by this statement?}
\vspace{3mm}
\\
A. Einstein wants to emphasize how useless classical physics is in order to support his new theory. He thus tries to create a dialogue suitable for this purpose.
\vspace{1mm}
\\
B. Einstein means to highlight that Newton’s physical laws are valid, but we do not have a real physical example to which we can directly apply them. We cannot completely isolate a body, and even if we could, is it legitimate to study its motion?
\vspace{1mm}
\\
C. Einstein means that the laws of classical physics are always valid, but they are applied to the wrong reference frames.
\item \textbf{3.}
\textit{In an inertial reference system, as described by the classical physicist, does gravity act?}
\vspace{3mm}
\\
A. Yes, of course, gravity is a force that is always present on Earth.
\vspace{1mm}
\\
B. No, because gravity is a force, and no forces should act on such a system.
\vspace{1mm}
\\
C. I don’t know. It certainly acts because it is always present on Earth, but we can pretend not to consider it.
\item \textbf{4.}
\textit{This text focuses on inertial reference systems. Why are they important? What does Einstein want to emphasize?}
\vspace{3mm}
\\
A. Inertial reference systems are systems in which Newton’s first law does not hold. Einstein wants to point out that the dynamics of the universe cannot be studied using classical physics.
\vspace{1mm}
\\
B. Inertial reference systems are systems in which the laws of physics always have the same mathematical form. This is also what Einstein wants to show.
\vspace{1mm}
\\
C. Inertial reference systems are systems in which Newton’s first law holds. Einstein emphasizes their definition to highlight that, in reality, as described by classical physics, no reference systems exist in which the laws of physics always have the same mathematical form.
\item \textbf{5.}
\textit{Who is right in the text?}
\vspace{3mm}
\\
A. Einstein
\vspace{1mm}
\\
B. The classical physicist
\vspace{1mm}
\\
C. Neither of them
\vspace{1mm}
\\
D. Both
\end{itemize}

\section{{Final Test}} \label{app:finaltest}
{\begin{itemize}
\item \textbf{1.}
\textit{According to Newton's proposed theory, gravity is an attractive force present between two bodies with mass. Einstein opposed this definition of gravity; according to him, gravity:}
\vspace{3mm}
\\
A. Is not a force.
\vspace{1mm}
\\
B. Is an attractive force but is exerted only by the body with greater mass on the body with lesser mass.
\vspace{1mm}
\\
C. Is an attractive force but is exerted only by the body with lesser mass on the body with greater mass.
\vspace{1mm}
\\
D. Is a force that arises from previous contact between the two bodies involved.
\item \textbf{2.}
\textit{Through the free-fall elevator experiment, we observed how an object inside it floats at the moment the elevator falls. This happens because:}
\vspace{3mm}
\\
A. We are in an inertial reference frame, and therefore gravity is not present. Since the gravitational force is no longer acting, the object's weight becomes 0 kg.
\vspace{1mm}
\\
B. The elevator's acceleration, which is g, causes us to experience an apparent force equal and opposite to the acceleration with which we are attracted to the Earth. These two forces cancel out, making the object float.
\vspace{1mm}
\\
C. The elevator accelerates, but the object inside does not; it remains "stationary" within the elevator, which is why it floats. In other words, the elevator moves relative to the object, but the object does not move.
\vspace{1mm}
\\
D. It is an issue with the experimental apparatus. The distance traveled by the free-falling elevator is very short, and the dynamometer inside oscillates. The fact that the recorded weight is zero is just a coincidence.
\item \textbf{3.}
\textit{We have stated Einstein's Equivalence Principle together. It is related to the fact that:}
\vspace{3mm}
\\
A. The Earth orbits the Sun.
\vspace{1mm}
\\
B. Staying still on Earth is equivalent to falling freely inside an elevator.
\vspace{1mm}
\\
C. Gravitational mass and inertial mass are equivalent.
\vspace{1mm}
\\
D. Floating in space and standing still on Earth are two equivalent situations.
\item \textbf{4.}
\textit{If a marble is launched horizontally inside a free-falling elevator, it will move only in the horizontal direction. How would you explain this phenomenon?}
\vspace{3mm}
\\
A. We are in an inertial reference frame, so gravity does not act, canceling the marble’s motion in the vertical direction.
\vspace{1mm}
\\
B. The motion of the marble is linked to the position of a guide placed horizontally inside the elevator, which makes it move horizontally.
\vspace{1mm}
\\
C. The marble, having mass, also experiences the acceleration g as it moves in free fall. As a result, an apparent force of double intensity and opposite to the force pulling us downward is created, canceling the marble’s motion in the vertical direction.
\vspace{1mm}
\\
D. The marble, having mass, also experiences the acceleration g as it moves in free fall. As a result, an apparent force equal and opposite to the force pulling us downward is created, canceling the marble’s motion in the vertical direction.
\vspace{5mm}
\\
Here are three questions related to the texts: "The Gaia Mission," "The Photon Sphere of Black Holes," and "The GPS." Answer them using the information from the texts and the concepts discussed in class.
\item \textbf{5.}
\textit{As explained in the first text, the Gaia mission aims to create a three-dimensional map of our galaxy. Referring to what we discussed in class, explain what potential challenges need to be considered and why General Relativity is essential for this mission.}
\vspace{3mm}
\\
\item \textbf{6.}
\textit{Why is it stated in the second text that light bends near a black hole? Explain in what sense we can say that light curves according to Newton and why it curves according to Einstein.}
\vspace{3mm}
\\
\item \textbf{7.}
\textit{How does the clock on a satellite remain synchronized with the clock on the device being used with GPS?}
\vspace{3mm}
\\
\end{itemize}}

\section{Texts}

\subsection{The Dialogue with a Classical Physicist} \label{app:dialogue}
In order to be more aware of this difficulty, let us interview the classical physicist and ask him some simple questions:\\
“What is an inertial system?”\\
“It is a Coordinate System (CS) in which the laws of mechanics are valid. A body on which no external forces are acting moves uniformly in such a CS. This property thus enables us to distinguish an inertial CS from any other.”\\
“But what does it mean to say that no forces are acting on a body?”\\
“It simply means that the body moves uniformly in an inertial CS.”\\
Here we could once more put the question: “What is an inertial CS?” But since there is little hope of obtaining an answer differing from the above, let us try to gain some concrete information by changing the question:\\
“Is a CS rigidly connected with the earth an inertial one?”\\
“No, because the laws of mechanics are not rigorously valid on the earth, due to its rotation. A CS rigidly connected with the sun can be regarded for many problems as an inertial CS; but when we speak of the rotating sun, we again understand that a CS connected with it cannot be regarded as strictly inertial.”\\
“Then what, concretely, is your inertial CS, and how is its state of motion to be chosen?”\\
“Tt is merely a useful fiction and I have no idea how to realize it.If I could only get far away from all material bodies and free myself from all external influences, my CS would then be inertial.”\\
“But what do you mean by a CS free from all external influences?”\\
“I mean that the CS is inertial.”\\
Once more we are back at our initial question! Our interview reveals a grave difficulty in classical physics. We have laws, but do not know what frame to refer them to, and our whole physical structure seems to be built on sand.\\

%\textit{From the book "The Evolution of Physics" \cite{einstein1966evolution}, Chapter III.}

\subsection{{The Gaia mission}} \label{app:gaia}
{Gaia is a mission of the ESA scientific program aimed at creating a three-dimensional map of our galaxy, revealing its composition, formation, and evolution.
\\
The launch took place on December 19, 2013, using the Soyuz-Fregat launcher, and the satellite was placed in orbit around L2.
\\
The mission will continuously scan the entire sky by leveraging the satellite's rotation and precession movements: each area of the sky will be observed approximately seventy times during the satellite's operational lifetime.
\\
The European scientific community's participation in the mission includes the responsibility of processing the vast amount of data that will be generated. This task is carried out by the Data Processing and Analysis Consortium (DPAC), a consortium of European research institutes established by European scientists in response to an ESA Announcement of Opportunity.}

\subsection{{The photon sphere of the black holes}} \label{app:bh}
{A photon sphere (or photonic sphere) is a spherical region of space where gravity is strong enough to force photons to move in orbital paths. The formula to determine the radius for a circular photon orbit is: $r=\frac{3GM}{c^{2}}$.\\
Due to this equation, photon spheres can only exist in the space surrounding an extremely compact object, such as a black hole.
\\
Since photons travel near the event horizon of a black hole, they can escape its gravitational pull by moving in an almost vertical direction, known as the escape cone. A photon at the boundary of this cone will not fully escape the black hole’s gravity; instead, it will orbit around it, although in unstable orbits.}

\subsection{{The GPS}} \label{app:gps}
{The Global Positioning System (GPS) is a U.S. military satellite-based positioning and navigation system.
\\
Using a dedicated network of artificial satellites in orbit, it provides a mobile terminal or GPS receiver with information about its geographic coordinates and time under any weather conditions, anywhere on Earth or in its immediate vicinity, as long as there is an unobstructed connection with at least four satellites of the system. Localization occurs through the transmission of a radio signal from each satellite and the processing of the received signals by the receiver.
\\
It is managed by the U.S. government and is freely accessible to anyone with a GPS receiver. Its current level of accuracy is within a few meters, depending on weather conditions, the availability and position of the satellites relative to the receiver, the quality and type of receiver, radio signal propagation effects in the ionosphere and troposphere (e.g., refraction), and relativistic effects.
\\
The clocks onboard the satellites are corrected for the effects predicted by relativity, which causes a time shift on the satellites. Observing this shift is considered a real-world confirmation of Einstein’s theory. The detected relativistic effect matches the theoretical expectation within measurement accuracy limits. The shift is the combined result of two factors: due to the satellite’s relative motion, its clock runs 7 microseconds per day slower than clocks on Earth, while due to the weaker gravitational potential at the satellite’s orbit, its clock runs 45 microseconds per day faster. The net result is that the satellite’s clock runs 38 microseconds per day ahead of clocks on Earth.
\\
To compensate for the time difference between satellite and Earth-based clocks, satellite clocks are electronically adjusted. Without these corrections, the GPS system would generate positioning errors of several kilometers per day and would not achieve the centimeter-level accuracy it is capable of.
\\
To reach the indicated accuracy, other synchronization errors must be considered, not just those of relativistic origin. For example, errors related to atmospheric signal propagation or onboard electronics delays can also affect accuracy. While relativistic errors are compensated, correcting atmospheric and electronic errors is more complex.}

%\begin{figure*}[h] 
%\appendix
%\section{Figures} \label{app:fig}
\begin{figure*}[h] 
\section{Figures} \label{app:fig}
\begin{minipage}{0.48\linewidth}
\centering 
\includegraphics[width=0.9\linewidth]{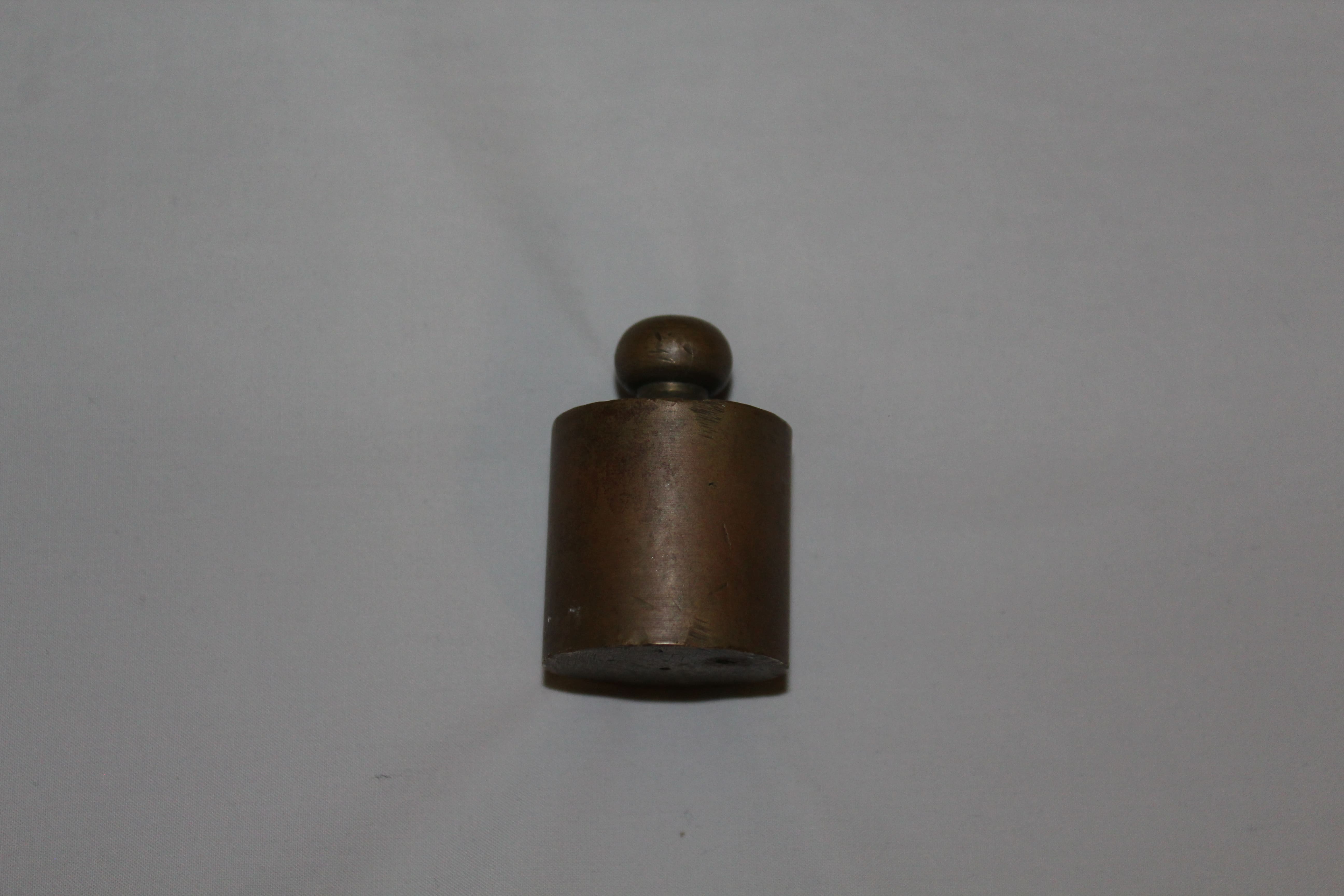} 
\caption{Weight lateral}\label{} 
\includegraphics[width=0.9\linewidth]{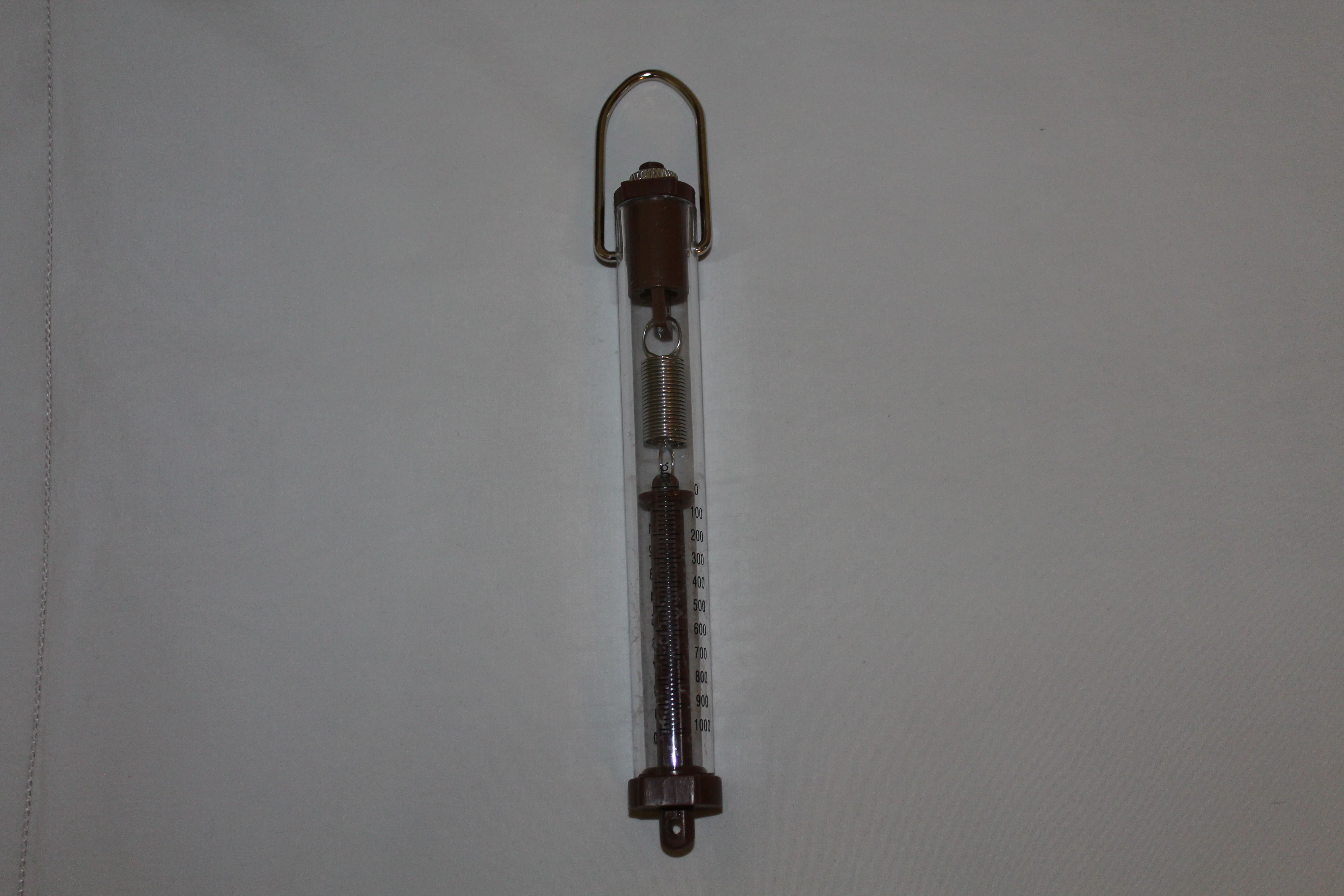} 
\caption{Dynamometer}\label{} 
\includegraphics[width=0.9\linewidth]{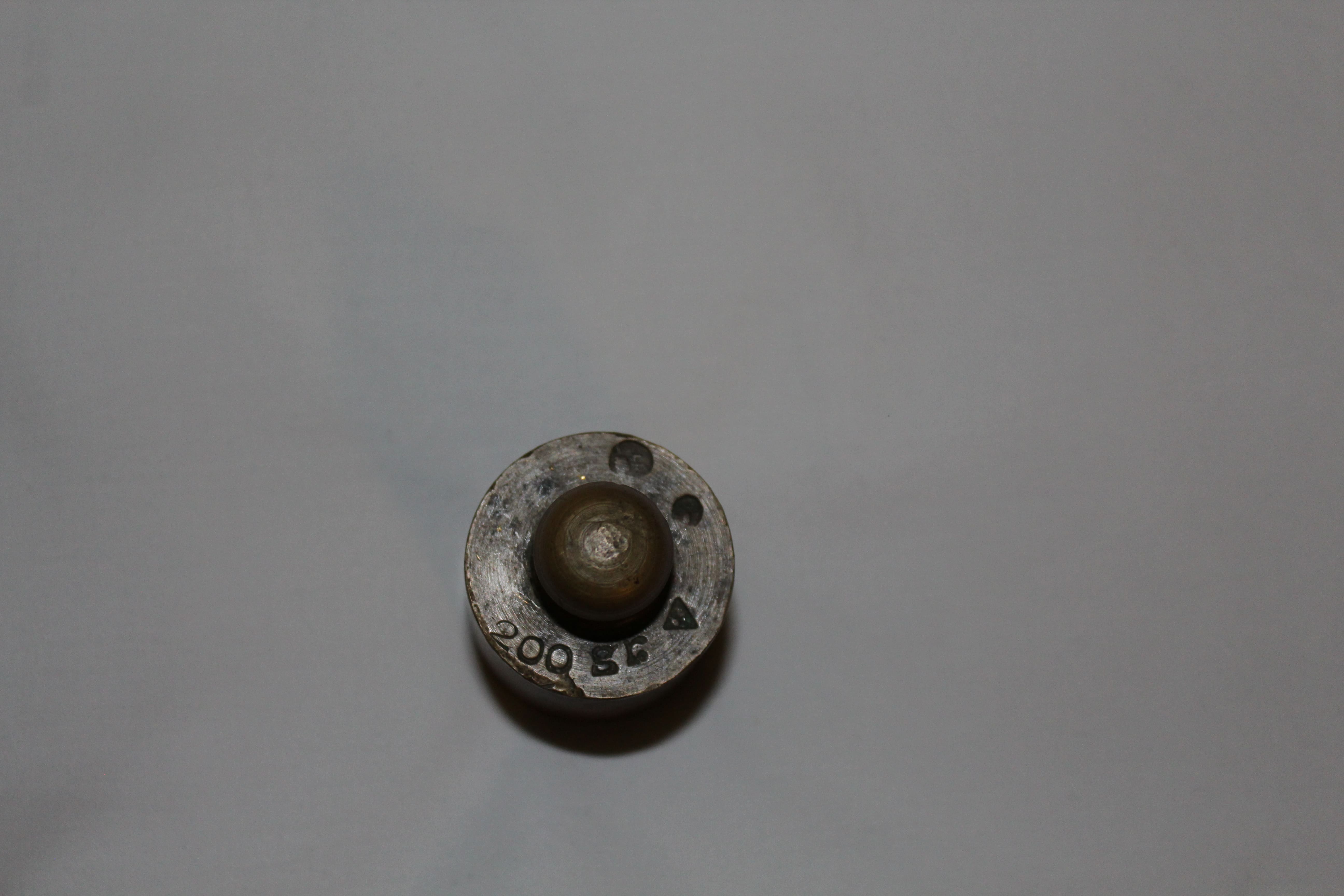} 
\caption{Weight frontal}\label{} 
\end{minipage}
\begin{minipage}{0.48\linewidth}
\centering
\includegraphics[width=0.9\linewidth]{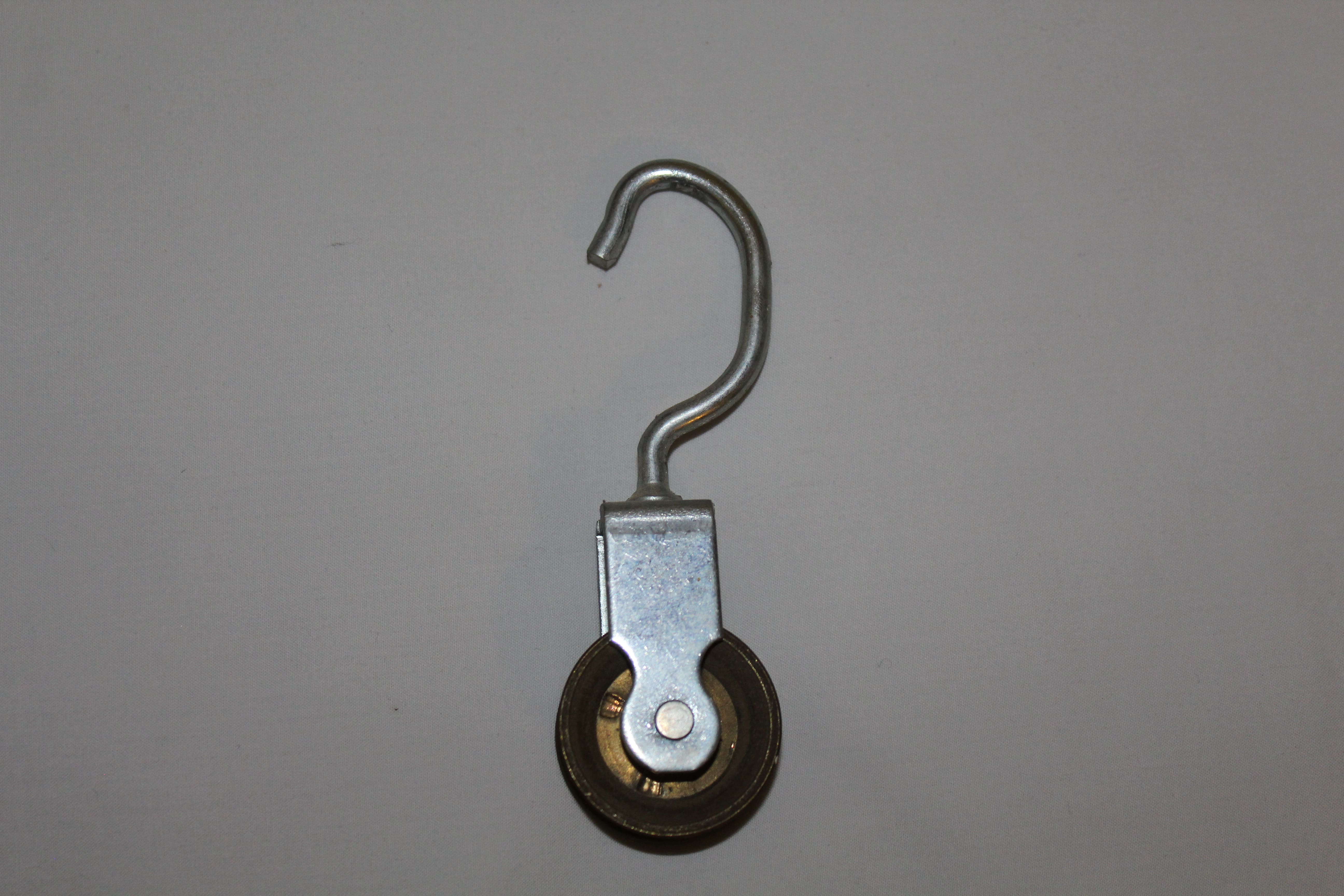} 
\caption{Pulley}\label{} 
\includegraphics[width=0.9\linewidth]{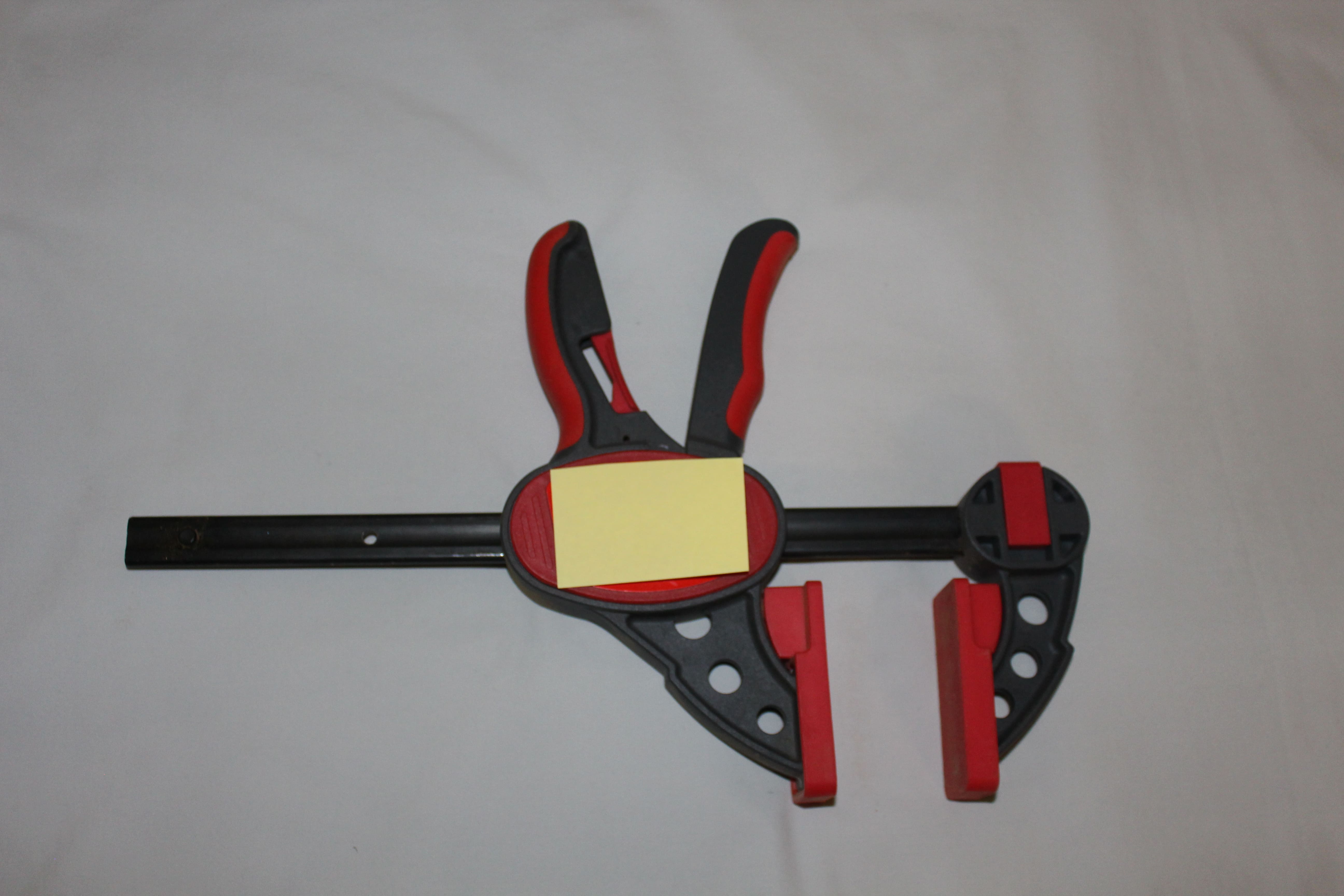} 
\caption{The central clamp}\label{} 
\centering 
\includegraphics[width=0.9\linewidth]{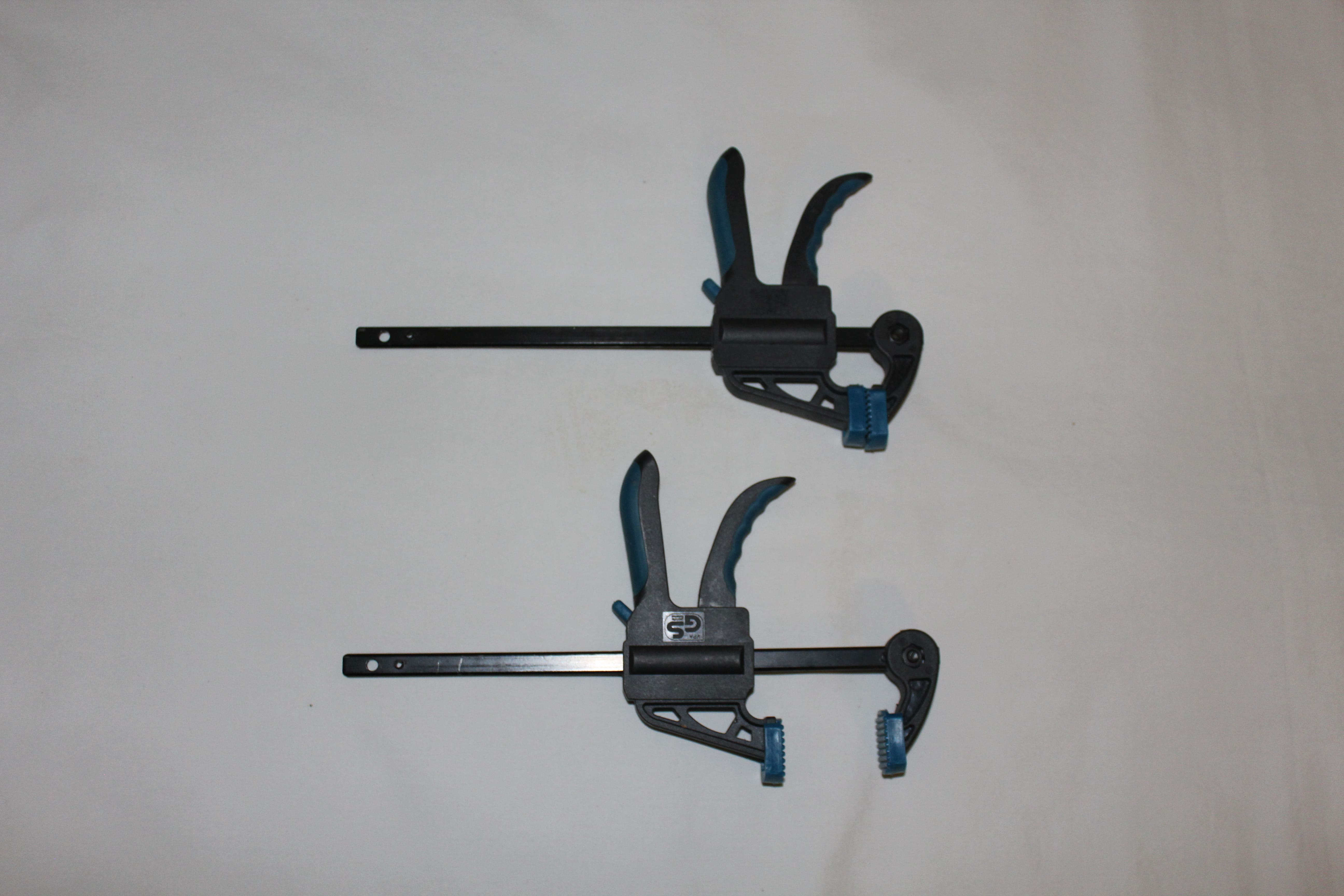} 
\caption{The lateral clamps}\label{} 
\end{minipage}
\end{figure*}

\begin{figure*}
\begin{minipage}{0.48\linewidth}
\centering
\includegraphics[width=0.9\linewidth]{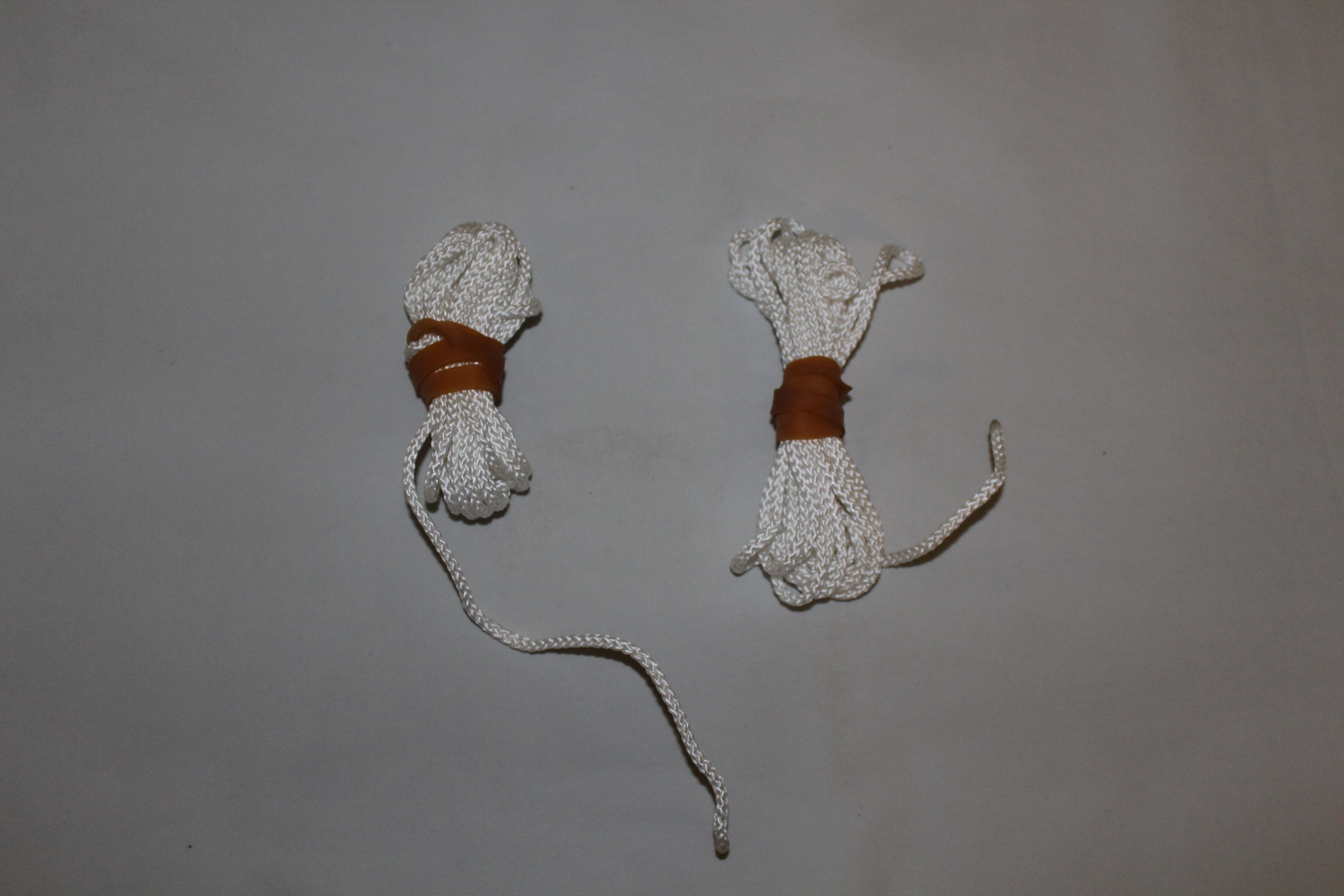} 
\caption{Ropes acting as guides}\label{} 
\includegraphics[width=0.9\linewidth]{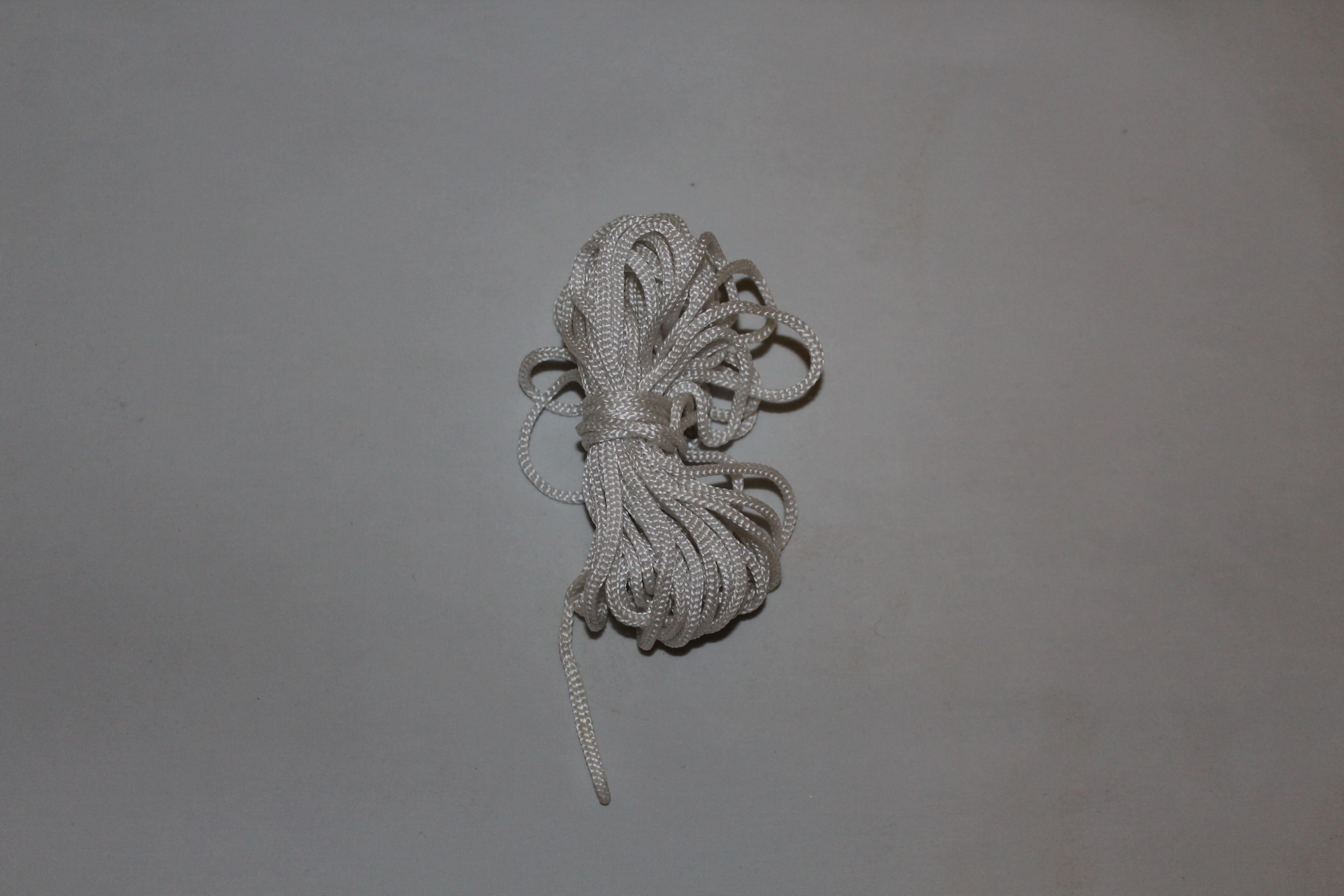} 
\caption{Rope lifting the box}\label{} 
\includegraphics[width=0.9\linewidth]{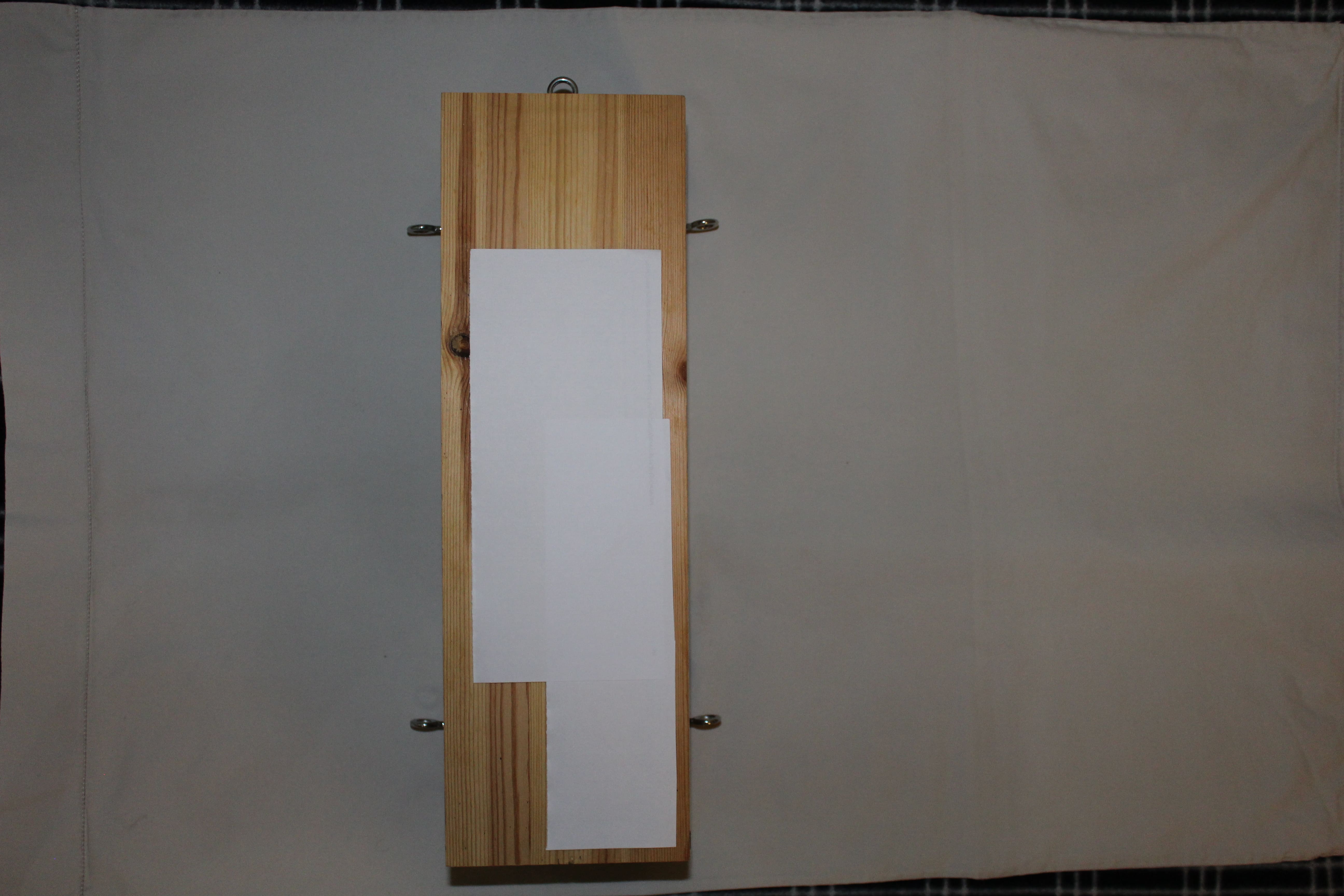} 
\caption{Closed box}\label{} 
\includegraphics[width=0.9\linewidth]{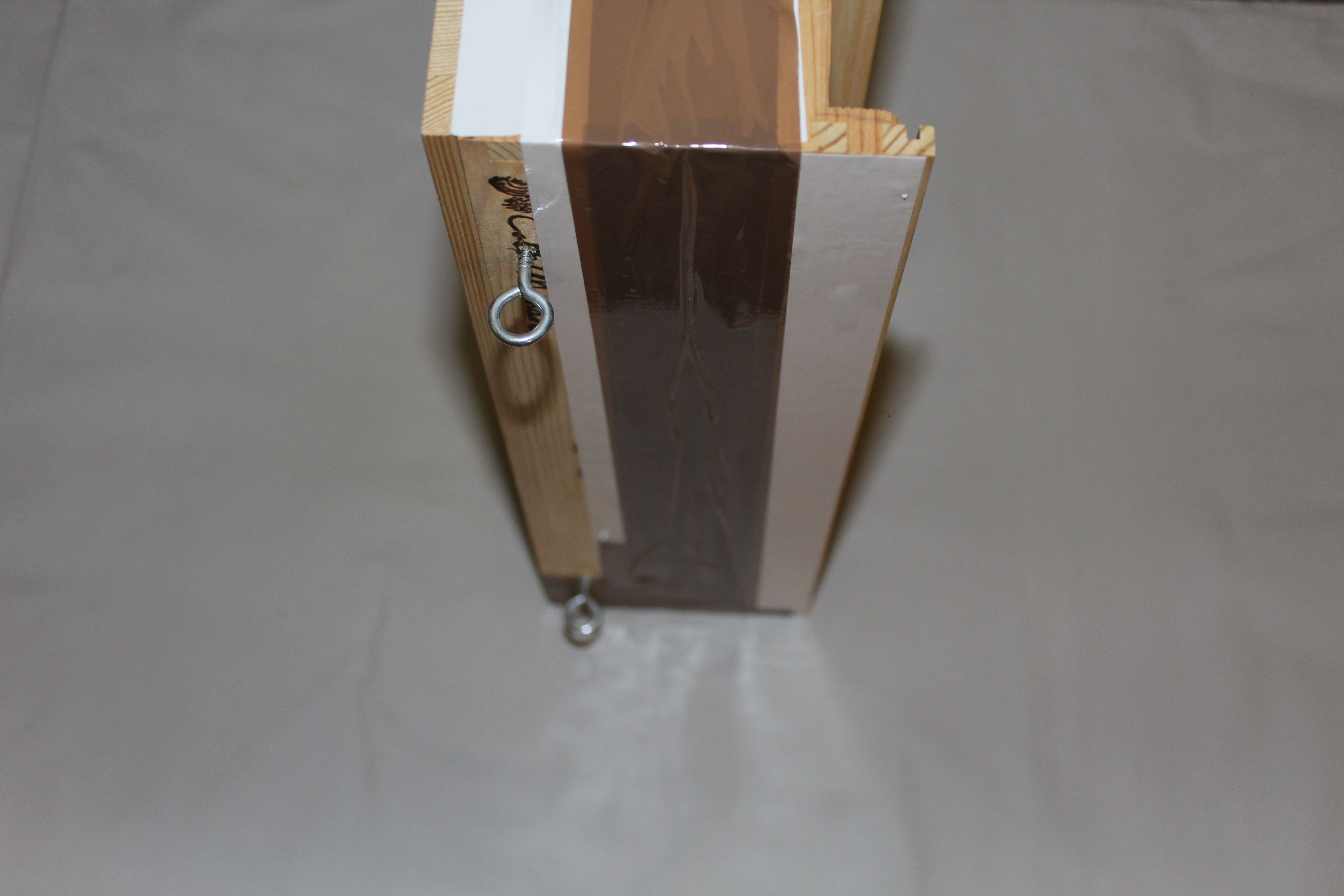} 
\caption{Box viewed from the side to show the two side rings}\label{} 
\end{minipage}
\begin{minipage}{0.48\linewidth}
\centering
\includegraphics[width=0.9\linewidth]{IMG_7715.pdf} 
\caption{Entire experimental apparatus}\label{} 
\end{minipage}
\end{figure*}

\end{document}